 \documentclass[preprint2]{aastex}
\shorttitle{Spitzer/IRAC and the Lambda Orionis cluster}
\shortauthors{Barrado y Navascu\'es et al.}
\begin{document}


\title{SPITZER: Accretion in Low Mass Stars and Brown Dwarfs in the 
Lambda Orionis Cluster\footnote{Based on observations collected Spitzer Space Telescope;
 at the German-Spanish Astronomical Center of Calar Alto
 jointly operated by the Max-Planck-Institut f\"ur Astronomie Heidelberg 
and the  Instituto de Astrof\'{\i}sica de Andaluc\'{\i}a (CSIC);
and at the WHT operated on the island of La Palma by the Isaac Newton Group in the Spanish Observatorio del
 Roque de los Muchachos of the Instituto de Astrofísica de Canarias} }
\author{David Barrado y Navascu\'es}
\affil{Laboratorio de Astrof\'{\i}sica Espacial y F\'{\i}sica Fundamental,
LAEFF-INTA, P.O. Box 50727, E-28080 Madrid, SPAIN}
\email{barrado@laeff.esa.es}

\author{John R. Stauffer}
\affil{Spitzer Science Center, California Institute of Technology, Pasadena, CA 91125}

\author{Mar\'{\i}a Morales-Calder\'on, Amelia Bayo}
\affil{Laboratorio de Astrof\'{\i}sica Espacial y F\'{\i}sica Fundamental, LAEFF-INTA, P.O. Box 50727, E-28080 Madrid, SPAIN}

\author{Giovanni Fazzio, Tom Megeath, Lori Allen}
\affil{Harvard Smithsonian Center for Astrophysics, Cambridge, MA 02138}

\author{Lee W. Hartmann, Nuria Calvet}
\affil{Department of Astronomy, University of Michigan}



\begin{abstract}
We present multi-wavelength optical and infrared photometry of 170
previously known low mass stars and brown dwarfs of the 5 Myr Collinder 69 cluster
(Lambda Orionis).  The new photometry supports cluster membership for
most of them, with less than 15\% of the previous candidates
identified as probable non-members.  The near
infrared photometry allows us to identify stars with IR excesses, and
we find that the Class~II population is very large, around 25\% for
stars (in the spectral range M0 - M6.5) and 40\% for brown dwarfs, down to
0.04 M$_\odot$, despite the fact that the H$\alpha$\
equivalent width is low for a significant fraction of them.  In addition, 
there are a number of substellar objects, classified as Class~III, that 
have optically thin disks. The Class~II 
members are distributed in an inhomogeneous way, lying preferentially
in a filament running toward the south-east.  The IR excesses for the
Collinder 69 members range from pure Class~II (flat or nearly flat
spectra longward of 1 $\mu$m), to transition disks with no near-IR
excess but excesses beginning within the IRAC wavelength range, to two
 stars with excess only detected at 24 $\mu$m.
Collinder 69 thus appears to be at an age where it provides a natural
laboratory for the study of primordial disks and their dissipation.

\end{abstract}





\newpage

\section{Introduction}

The star-formation process appears to operate successfully over a wide
range of initial conditions.  In regions like Taurus, groups of a few
stars to a few tens of stars are the norm.  The molecular gas in
Taurus is arranged in a number of nearly parallel filaments, possibly
aligned with the local magnetic field, and with the small stellar
groups sited near end-points of the filaments \citep{Hartmann04IAU}.
No high mass stars have been formed in the Taurus groups, and the
Initial Mass Function (hereafter, IMF) also appears to be relatively
deficient in brown dwarfs \citep{Briceno03IAU,Luhman04} --but see also
\citep{Guieu06} for an alternate view. The Taurus groups are not
gravitationally bound, and will disperse into the field on short
timescales. At the other end of the mass spectrum, regions like the
Trapezium cluster and its surrounding Orion Nebula cluster (ONC) have
produced hundreds of stars.  The ONC includes several O stars, with
the earliest having spectral type O6 and an estimated mass of order 35
M$_\odot$.  The very high stellar density in the ONC (10,000
stars/pc$^{3}$ at its center \citep{McCaughrean94}) suggests that
star-formation in the ONC was gravity dominated rather than magnetic
field dominated.  It is uncertain whether the ONC is currently
gravitationally bound or not, but it is presumably at least regions
like the ONC that are the progenitors of long-lived open clusters like
the Pleiades.  UV photoionization and ablation from O star winds
likely acts to truncate the circumstellar disks of low mass stars in
the ONC, with potential consequences for giant planet formation.

An interesting intermediate scale of star-formation is represented by
the Lambda Ori association.  The central cluster in the association
--normally designated as Collinder 69 or the Lambda Orionis cluster--
includes at least one O star, the eponymous $\lambda$ Ori, with
spectral type O8III.  However, a number of lines of evidence suggests
that one of the Coll 69 stars has already passed through its post-main
sequence evolution and become a supernova, and hence indicating it was
more massive than Lambda Ori (see the complete Initial Mass Function in 
\citet{Barrado2005a}).  A census of the stars in Coll 69 by
\citet{Dolan01} --hereafter, DM-- indicates that the cluster is now
strongly unbound.  DM argue that this is due to rapid removal of
molecular gas from the region that occurred about 1 Myr ago when the
supernova exploded.  They interpreted the color-magnitude diagram of
Coll 69 as indicating a significant age spread with a maximum age of
order 6 Myr; an alternative interpretation is that the cluster has
negligible age spread (with age $\sim$6 Myr) and a significant number
of binary stars.  While DM identified a large population of low mass
stars in Coll 69, only four of 72 for which they obtained spectra are
classical T Tauri stars (based on their H$\alpha$\ emission equivalent
widths).  Much younger stars, including classical T Tauri stars, are
present elsewhere in the Lambda Ori SFR, which DM attribute to
star-formation triggered by the supernova remnant shock wave impacting
pre-existing molecular cores in the region (the Barnard 30 and
Barnard 35 dark clouds, in particular).

We have obtained Spitzer space telescope IRAC and MIPS imaging of a
$\sim$one square degree region centered on the star $\lambda$ Ori in
order to (a) search for circumstellar disks of members of the Coll 69
cluster and (b) attempt to identify new, very low mass members of the
cluster in order to determine better the cluster IMF 
(in a forthcoming paper).
  In \S2, we
describe the new observations we have obtained; and in \S3 we use
those data to reconsider cluster membership.  In \S4 we use the new
candidate member list and the IR photometry to determine the fraction
of cluster members with circumstellar disks in both the stellar and
substellar domain, and we sort the  stars with disks according to their
spectral slope from 1 to 24 $\mu$m.

\section{The data}

\subsection{Optical and Near Infrared photometry}

The optical and the near IR data for the bright candidate members come
from \citet{Barrado04} --hereafter, Paper~I.  The $RI$ --Cousins
system-- data were collected with the CFHT in 1999, whereas the $JHKs$
come from the 2MASS All Sky Survey \citep{Cutri03}.  For cluster
members, the completeness limit is located at $I$(complete,cluster)
$\sim$ 20.2 mag, whereas 2MASS provides near infrared data down to a
limiting magnitude of $J$=16.8, $H$=16.5, and $Ks$=15.7 mag.  In some
cases, low resolution spectroscopy in the optical, which provides
spectral types and H$\alpha$\ equivalent widths, is also available.
Twenty-five objects out of the 170 CFHT1999 candidate members are in
common with \citet{Dolan99,Dolan01}. Those 25 stars also have
H$\alpha$\ and lithium equivalent widths, and radial velocities.

\subsubsection{New deep Near Infrared photometry}

For the objects with large error in 2MASS JHKs, or without this type
of data due to their intrinsic faintness, we have obtained additional
measurements with the WHT (La Palma Observatory, Spain) and INGRID
(4.1$\times$4.1 arcmin FOV) in November 2002 and February 2003, and
with the Calar Alto 3.5m telescope (Almeria, Spain) and Omega2000 in
October 2005 (15.36$\times$15.36 arcmin FOV).  In all cases, for each
position, we took five individual exposures of 60 seconds each, with 
small offsets of a few arcseconds, thus totalling 5 minutes.  In the case
of the campaigns with INGRID, we observed the area around the star
$\lambda$ Orionis creating a grid. Essentially, we have observed about
2/3 of the CFHT 1999 optical survey region in J (in the area around the star
and west of it), with some coverage in H and K.  On the other hand,
the Omega2000 observations, taken under a Director's Discretionary Time
program, were focused on the faint candidate members.  Except for one
object (LOri154), we collected observations in the J, H and Ks
filters.  The conditions of the first observing run with INGRID were
photometric, and we calibrated the data using standard stars from
\citet{Hunt98} observed throughout the nights of the run. The average
seeing was 0.9 arcsec.  We had cloud cover during the second run with
INGRID, and the data were calibrated using the 2MASS catalog and the
stars present in each individual image. The dispersion of this
calibration is $\sigma$=0.05 magnitudes in each filter, with a seeing
of about 1.0 arcsec. Finally, no standard stars were observed during
the DDT observations at Calar Alto. The seeing in this case was 1.2
arcsec.  The faint Lambda Orionis candidate members were calibrated
using also 2MASS data. In this last case, the dispersion is somewhat
higher, probably due to the worse seeing and the larger angular pixel
scale of the detector, with $\sigma$=0.1 mag. Note that this is
dispersion not the error in the calibration. These values correspond
to the FWHM of the gaussian distribution of the values
zeropoint(i)=mag$_{raw}$(i)-mag$_{2MASS}$(i), for any star $i$, which
also includes the photometric errors in the 2MASS photometry and any
contribution due to the cluster stars being photometrically variable.
Since there is a large number of stars per field (up to 1,000 in the
Omega2000 images), the peak of this distribution can be easily
identified and the zero points derived. A better estimate of the error
in the calibration is based on the distance between mean, median and
mode values, which are smaller than half of the FWHM (in the case of
the mean and the median, almost identical to the hundredth of
magnitude).  Therefore, the errors in the calibration can be estimated
as 0.025 and 0.05 magnitudes for the INGRID and the Omega2000
datasets, respectively. 

All the data were processed and analyzed with IRAF\footnote{IRAF is
distributed by National Optical Astronomy Observatories, which is
operated by the Association of Universities for Research in Astronomy,
Inc., under contract to the National Science Foundation, USA}, using
aperture photometry. These measurements, for 166 candidate members,  are listed in
 Table 1 (WHT/INGRID) and Table 3 (CAHA/Omega2000). Note
that the errors listed in the table correspond to the values produced
by the $phot$ task with the $digi.apphot$ package and does not include the errors
in the calibrations.

\subsection{Spitzer imaging}

Our Spitzer data were collected during March 15 (MIPS) and October 11
(IRAC), 2004, as part of a GTO program.  The InfraRed Array
Camera (IRAC, \citet{Fazio04}) is a four channel camera which 
takes images at 3.6, 4.5, 5.8, and 8.0~$\mu$m with a field of view
that covers $\sim$5.2$\times$5.2~arcmin. IRAC imaging was performed
in mapping mode with individual exposures of 12 seconds ``frametime"
(corresponding to 10.4 second exposure times) and three dithers at
each map step.  In order to keep the total observation time for a
given map under three hours, the Lambda Ori map was broken into two
segments, each of size 28.75$\times$61.5 arcmin - one offset west of
the star $\lambda$ Ori and the other offset to the east, with the
combined image covering an area of 57$\times$61.5 arcmin, leaving the
star $\lambda$ Orionis approximately at the center. Each of the
IRAC images from the Spitzer Science Center pipeline were corrected
for instrumental artifacts using an IDL routine developed by S. Carey
and then combined into the mosaics at each of the four bandpasses
using the MOPEX package \citep{Makovoz05}. Note that the IRAC images
do not cover exactly the same FOV in all bands, providing a slice
north of the star with data at 3.6 and 5.8 micron, and another slice
south of it with photometry at 4.5 and 8.0 microns.  The size of these
strips are about 57$\times$6.7 arcmin in both cases.  The 
Multiband Imaging Photometer for Spitzer (MIPS, \citet{Rieke04}) was
used to map the cluster with a medium rate scan mode and 12 legs
separated by 302 arcsec in the cross scan direction.  The total
effective integration time per point on the sky at 24~$\mu$m for most
points in the map was 40 seconds, and the mosaic covered an area
of 60.5$\times$98.75 arcmin centered around the star $\lambda$
Orionis. Since there were no visible artifacts in the pipeline
mosaics for MIPS 24~$\mu$m we used them as our starting point to
extract the photometry. We obtained MIPS 70 $\mu$m and 160 $\mu$m
imaging of the $\lambda$ Ori region, but very few point sources were
detected and we do not report those data in this paper.

The analysis of the data was done with IRAF. First, we detected
objects in each image using the ``starfind'' command. Since the images
in the [3.6] and [4.5] bands are deeper than those in the [5.8] and
[8.0] bands, and since the fluxes of most objects are brighter at
those wavelengths, the number of detections are much larger at the
IRAC short wavelengths than at the longer ones. Only a relatively few
objects have been detected at 24~$\mu$m with MIPS.
As a summary, 164 objects were detected at at 3.6 and 4.5 micron, 
145 at 5.8 micron, 139 at 8.0 micron  
and  13 at 24 micron.

We have performed aperture photometry to derive fluxes for C69
cluster members.  For the IRAC
mosaics we used an aperture of 4 pixels radius, and the sky was computed
using a circular annulus 4 pixels wide, starting at a radius 4 pixels
away from the center. It is necessary to apply an aperture correction
to our 4-pixel aperture photometry in order to estimate the flux
for a 10-pixel aperture, because the latter is the aperture size used
to determine the IRAC flux calibration.  In some cases, due to the presence
of nearby stars, hot pixels, or because of their faintness, a 2 pixel
aperture and the appropriate aperture correction were used (see notes to Table 3). 
 For the MIPS
photometry at 24~ $\mu$m, we used a 5.31 pixels (13 arcsec) aperture
and a sky annulus from 8.16~pixels (20 arcsec) to 13.06~pixels (32
arcsec). An aperture correction was also applied. Table 2 provides the
zero points, aperture corrections and conversion factors between
magnitudes and Jansky, as provided by the Spitzer Science Center website.


\subsubsection{Data cross-correlation}

The coverage on the sky of our Spitzer/IRAC data is an approximate
square of about 1 sq.deg, centered on the star $\lambda$ Orionis. The
optical data taken with the CFHT in 1999 covers an area of
42$\times$28 arcmin, again leaving the star in the center of this
rectangle.  Therefore, the optical survey is completely included in
the Spitzer mapping, and we have been able to look for the counterpart
of the cluster candidates presented in Paper~I. The analysis of the
area covered by Spitzer but without optical imaging in the CFHT1999
survey will be discussed in a forthcoming paper. We have not been able
to obtain reliable Spitzer photometry for some candidate members from Paper I,
especially at the faint end of the cluster sequence. The faintest detected object, 
LOri167, depending on the isochrone and the model, would have a mass of $\sim$0.017
 $M_\odot$, if it is a member (\citet{Barrado2005b}).

The results are listed in Table~3, where non-members and members are
included, respectively (see next section for the discussion about the
membership). In both cases, we include data corresponding to the bands
R and I --from CFHT--, J, H and Ks --from 2MASS and  CAHA--,
[3.6], [4.5], [5.8], and [8.0] --from IRAC-- and [24] --from MIPS.
Additional near IR photometry from WHT can be found in Table~1.


\section{Color-Color and Color-Magnitude Diagrams and new membership assignment}

Before discussing membership of the Paper I stars based on all of the
new optical and IR data, we have made an initial selection based on
the IRAC colors. Figure 1 (see further discussion in the next section)
displays a Color-Color Diagram with the four IRAC bands. We have found
that 31 objects fall in the area defined by \citet{Allen04} and \citet{Megeath2004} as Class~II 
objects (ie, TTauri stars). Another two candidate members are
located in the region corresponding to Class~I/II objects. We consider
all these 33 objects as bona-fide members of the C69 cluster.
\citet{Harvey06} have discussed the confusion by extragalactic and other 
sources when analysing Spitzer data (in their case, they used Serpens, a cloud
having a large extinction). 
We believe that this contribution is negligible for our Lambda Ori
data, since those Class II objects detected at 24 micron are in the TTauri area
 defined by \citet{Sicilia05}, as displayed in her figure 5.
 There can be a higher level of contamination among the objects classiefied as Class~III.
All of the objects in Figure 1 had previously been identified as
cluster candidate members based on optical CMDs --it is unlikely that a significant
number of AGN would have passed both our optical and our IR criteria (and also have
 been unresolved in our optical CFHT images). Moreover, prior to our Spitzer data,
 only 25\% candidate members which had optical, near-IR data and 
 optical low-resolution spectroscopy turned out to be non-members
(Paper I). After adding the Spitzer photometry,
we are quite confident in the membership of the selection.

Figure~2 and Figure~3 display several color-magnitude diagrams
(CMD) using the data listed in Tables~1 and 3. In the case of the
panels of the first figure, we present optical and IR, including
Spitzer/IRAC data; whereas in the second set of figures only IR data
are plotted. For the sake of simplicity, we have also removed the
non-members from Figure~3.

Based on these diagrams and on the spectroscopic information included
in Paper~I, we have reclassified the candidate members as belonging or
not to the cluster.   In color-magnitude diagrams, C69 members lie in
a fairly well-defined locus, 
with a lower bound that coincides approximately with the 20 Myr
 isochrone in this particular set
of theoretical models (Baraffe et al. 1998).
  Stars that fall well below (or blueward)
of that locus are likely non-members; stars that fall above or redward
of that locus are retained because they could have IR excesses or above
average reddening.   We combine the ``votes" from several CMD's to
yield a qualitative membership determination, essentially yes, no or
maybe.  In total, out of 170 candidates, 19 are probable non-members, four
have dubious membership and the rest (147 objects) seem to be bona-fide
 members of the cluster.  Therefore, the ratio of non-members
to initial candidate members is 13.5 \%. 
In any case, only additional spectroscopy (particularly medium and high resolution) 
can be used to establish the real status. Proper motion might be helpful, 
but as shown by Bouy et al. (2007), some bona-fide member 
can appear to have discrepant proper motions
 when compared with the average values of the association.
Table~4 shows
the results for each candidate in the different tests used to
determine its membership, the membership as in Paper~I, and the final
membership based on the new photometry. The second and last columns show
the spectroscopic information. Note that the degree of confidence in
the new membership classification varies depending on the available
information and in any event it is always a matter of probability.
 
As Table~3 shows, the Spitzer/IRAC data does not match completely the
limiting magnitudes of either our optical survey with CFHT or the
2MASS $JHKs$ data.  In the case of the band [3.6], essentially all the
Lambda Orionis candidate members should have been detected (except
perhaps the faintest ones, at about $I$=22 mag). Some objects in the
faint end have 3.6 micron data, but lack 2MASS NIR, although in most
cases we have supplied it with our own deep NIR survey. In the case of
the Spitzer data at 4.5 micron, some additional candidate members
fainter than $I$=20.9 mag were not found, due to the limiting
magnitude of about [4.5]$_{lim}$=16.3 mag. The data at 8.0 micron only
reach [8.0]$_{lim}$=14.0 mag, which means that only cluster members
with about $I$=18.6 mag --or $Ks$=14.9-- can be detected at that
wavelength. This is important when discussing both the membership
status based on color-magnitude diagrams and the presence of infrared
excesses by examining color-color diagrams. Note, however, that
objects with IR excesses have fainter optical/near-IR counterparts
than predicted in the table.

Figure 4 presents another CMD with the optical magnitudes from the
CFHT survey ($R$ and $I$), where we display the 170 candidate members
using different symbols to distinguish their actual membership
status. Small dots correspond to non-members based on the previous
discussion, whereas plus symbols, crosses and circles denote probable
members. In the first case --in most cases due to their faintness--
they do not have a complete set of IRAC magnitudes, although they can
have either a measurement at 3.6 and 4.5, or even at 5.8 microns. In
the case of the objects represented by crosses, they have been
classified as Class~III objects (Weak-line TTauri stars and substellar
analogs if they indeed belong to the cluster) based on an IRAC
color-color diagram (see next section and Figure~1).  Finally, big
circles correspond to Class~II sources. The pollution rate seems to be
negligible in the magnitude range $I$=12-16 (1.2--0.17 M$_\odot$
approx, equivalent to M0 and M5, respectively), 
where our initial selection based on the optical and the near
infrared (2MASS data) has worked nicely. However, for fainter
candidates, the number of spurious members is very large and the
pollution rate amounts to about 15\% for objects with 16 $<$
$I$\ $<$ 19, and about 45\% for I$\ge$19 (approximately the magnitude
beyond the reach of the 2MASS survey).

At a distance of 400 pc and for an age of 5 Myr, and according to the
models by \citet{Baraffe98}, the substellar borderline is located at
about I=17.5 mag. Table~5 lists other values for different ages, as
well other bands --$J$, $Ks$ and $L'$-- discussed in this paper.
Among our 170 CFHT candidate members, there are 25 objects fainter
than that magnitude, and which pass all of our membership criteria
which are probable brown dwarfs.  Out of these 25 objects, 12 have
low-resolution spectroscopy and seem to be bona-fide members and,
therefore, brown dwarfs.  The other 13 objects are waiting for spectral
confirmation of their status.  Assuming an age of 3 or 8 Myr would
increase or decrease the number of brown dwarf by seven in each
case. In the first case (3 Myr), five out of the seven possible BDs
have spectroscopic membership, whereas in the second case only three
were observed in Paper~I. As a summary, we have found between 18
and 32 good brown dwarfs candidates (depending on the final age) in
the Lambda Orionis cluster, and between 17 and 9 have their nature
confirmed via low-resolution spectroscopy. Note that even this
technique does not preclude the possibility that a few among them
would actually be non-members.

Finally, the planetary mass domain starts at about $Ic$=21.5, using a
5 Myr isochrone (DUSTY models from \citet{Baraffe02}). In that region,
there is only one promising planetary mass candidate, LOri167
(\citet{Barrado2007AALetters}).
  
 
\section{Discussion}

\subsection{The Color-Color diagrams, the diagnostic of IR excess and the disk ratio}

The Spitzer/IRAC colors are a powerful tool to reveal the dust and,
therefore, the population of Class~I and II sources in a stellar
association. Figure 1 (after \citet{Allen04} and \citet{Megeath2004}) displays the colors
derived from the measurements at 3.6 minus 4.5 microns, versus those
obtained at 5.8 minus 8.0 microns. This diagram produces an excellent
diagnostic, allowing an easy discrimination between objects with and
without disks. Note that due to the limiting magnitudes of the IRAC
bands (see the discussion in previous section), objects fainter than
about $I$=18.6 mag cannot have a complete set of IRAC colors and
therefore cannot be plotted in the diagram. This fact imposes a limit
on our ability to discover mid-IR excesses at the faint end of the
cluster sequence.  For Lambda Orionis cluster members, assuming a
distance of 400 pc and an age of 5 Myr (and according to the models by
\citet{Baraffe98}), this limit is located at 0.040 M$_\odot$. Figure 1  
contains a substantial number of objects in the region corresponding
to the Class~II sources. In total, there are 31 objects located within
the solid rectangle out of 134 Lambda Orionis members with data in the
four IRAC bands. 
Among them, three (LOri045, LOri082 and LOri092) possibly have
relatively large photometric errors in their 5.8 $\mu$m flux, because
inspection of
their SEDs indicates they are likely diskless.
Two additional objects, LOri038 and LOri063, have
IRAC colors indicating Class~I/II (actually, LOri038 is very close to
the Class~II region). The SED (see below) indicates that both are Class~II stars.  
Therefore, the fraction of cluster members that are Class~II PMS stars 
based on their IR excesses is $\sim$22--25\%, for the spectral range M0--M6.5.  
This is different from what was inferred by
\citet{Dolan99, Dolan01} and by us (Paper~I), based on the
distribution of the H$\alpha$\ emission and near-infrared color-color
diagrams.  The Spitzer/IRAC data clearly demonstrate that Lambda
Orionis cluster does contain a significant number of stars with dusty
circumstellar disks. No embedded objects (Class~I) seem to be present,
in agreement with the age range for the association (3-8 Myr or even
slightly larger).  Note that our different IR excess frequency compared
to Dolan \& Mathieu may result from their sample being primarily of
higher mass stars than ours.

Figure~5 is a blow-up of the region in Figure 1 corresponding to the
Class~II sources.  We have also added big minus and plus symbols, and
large squares, to indicate those objects with measured H$\alpha$\
equivalent widths (in low- and medium-resolution spectrum). We have
used the saturation criterion by
\citet{Barrado03} to distinguish between objects with high W(H$\alpha$)
--plus symbols-- and normal W(H$\alpha$) --minus symbols. In
principal, an object with a W(H$\alpha$) value above the saturation
criterion is either accreting or is undergoing a flare episode. There
are two low mass stars (LOri050 and LOri063) with an H$\alpha$\
line broader than 200 km/s \citep{Muzerolle03}, another independent
indication of accretion (based on \citet{Natta04}, they should have
very large accretion rates $\sim$10$^{-9}$ $M_\odot$/yr).  The
theoretical disk models used to interpret IRAC Color-Color Diagram by
\citet{Allen04} suggest that the accretion rates increase from the
bottom-left to the top-right of the figure.  This is in agreement with
our results, since most of the accreting objects (assuming that strong
H$\alpha$\ is a good indicator of accretion) lie in the area of the
figure with the larger excesses (top-right). A couple of objects with very low
H$\alpha$\ emission are located near the edge of the Class~II
 area (bottom-left), a fact that suggest that they may have a
relatively thin disk, with small or negligible accretion. 
Actually these objects are surrounded by thin disks instead of 
thick primordial disks (see next section).

Regarding the brown dwarfs in the cluster, several probable members
(LOri126, LOri129, LOri131, LOri132, LOri139 and LOri140) are located within the
precinct of Classical TTauri stars. They are just at the border
between stars and substellar objects, with magnitudes in the range
$I$=17.52--18.21 and $J$=15.38--16.16 (the boundary is located at
$I$=17.55 and $J$=15.36 for 5~Myr, see Table~5).  In Paper~I we
presented low-resolution spectroscopy of LOri126, LOri139 and LOri140,
which suggests they are cluster members (the spectral types are M6.5,
M6.0 and M7.0 with a H$\alpha$\ equivalent width of 26.2, 19.7 and
72.8 \AA, respectively). In addition we have confirmed the
membership of LOri129 via medium-resolution spectroscopy (spectral
type, M6.0 with a H$\alpha$\ equivalent width of 12.1 \AA).
%

In total, there are 15 brown dwarf candidates with a complete set of
IRAC colors, six of which fall in the Class~II region, 
thus making the fraction of brown dwarfs with IR colors
indicative of circumstellar disks close to 40\% (down to 0.04 M$_\odot$),
 similar to the 50\% obtained by \citep{Bouy2007}
 in  Upper Sco brown dwarfs, 
 using mid-IR photometry or the 50\% derived by Guieu et al. (2006) in Taurus brown dwarfs 
with Spitzer.



\subsection{The Spectral Energy Distribution}

We have plotted the SEDs of our Lambda
Orionis candidate members in Figures~6-8. There is clearly a range
from approximately flat spectrum, to black-body in the near-IR but
starting to show excesses at IRAC wavelengths, to only showing excess
at 24 micron. A way to study the presence of a circumstellar disk around
an object is to analyze the shape of the SED.
 After \citet{Lada06} we have used the
3.6--8.0~$\mu$m slope for each source detected in at least three IRAC
bands to distinguish between objects with optically thick,
 primordial disks, objects surrounded by optically thin or anemic disks 
 and objects without disks.
The results of this test are presented in
Table~4. In Figures~6-8 the SEDs are sorted in agreement with their IRAC 
slope classification: 
diskless objects (slope index or $\alpha$$<$$-2.56$) in Figure 6,
 thick disks ($\alpha$$>$$-1.8$) in  Figure 7, 
and objects surrounded by thin disks 
($-1.8$$<$$\alpha$$<$$-2.56$) in Figure 8.
In this last figure we also include two low mass stars which present an excess 
only at 24 micron, due to a transition disk (see below).
 According to the IRAC  slope the fraction of cluster members
detected in at least three of the IRAC bands with optically thick
disks is 14\%, while the total disk fraction is found to be 31\% 
(similar to the 25\% derived with the IRAC CCD). This
fraction is lower than the 50\% found by \citet{Lada06} in IC348
(1-3~Myr) as expected due to the different age of the clusters.

Figure 9 illustrates the evolution of the disk fraction with the age
for several stellar associations (assuming that the infrared excess
serves as a proxy of the presence of a circumstellar disk). The ratios for the
 different associations come from IRAC data \citep{Hartmann05,Lada06,Sicilia06}
 in order to avoid different results depending on the technique used \citep{Bouy2007}.
The ratio for the Lambda Orionis cluster (Collinder 69) is about 30\% and,  as stated
before, for the objects below the substellar borderline, the fraction
of Classical TTauri objects seems to be larger. According to its older age, 
the thick disk fraction in Collinder~69 is lower than that of IC348 
(this fraction is represented by open squares in Figure~9).

Among the objects classified as Class~III sources from Figure 1, only
two (LOri043 and LOri065) have a measurement at [24] with an unambiguous
detection.  These two stars do not have excesses at 3.6 or 4.5 micron.
Therefore, they can be classified as transition objects, the
evolutionary link between the primordial disks and debris disks.
A third of the Class~II sources (11 out of 33) have measurements in
the [24] band, all of them with clear excess, as expected from
their Class~II status. The
lack of IR excesses at shorter wavelengths for LOri043 and LOri065
probably stems from an inner disk hole or at least less inner dust
than for the Class~II sources.  Models of similar 24 $\mu$m-only 
excess sources and a discussion of their disk-evolutionary significance
can be found in \citet{Sicilia06,Muzerolle06,DAlessio06}. 
Figure~10 shows the same diagram as in Figure~1 but the MIPS 24~$\mu$m
 information is included as dashed squares.  The small circles stand 
for objects having optically thin (dashed) or optically thick (solid) 
disks based on their IRAC slope. The diagram shows a smooth transition
 between the three types of objects: diskless, thin, and primordial 
disks. 
 LOri103  has a thin disk based on its 3.6--5.8~$\mu$m
 slope. It has been classified as Class~III due to its magnitude at 
8.0~$\mu$m but we believe that it is actually a Class~II source and the faint
magnitude at this  bandpass is probably due to the presence
 of a nebulosity.

There are some objects classified as Class~III sources by the 
IRAC CCD (they are outside, but  close to the Class~II area in the 
diagram), but have disks based on their IRAC slope. All these objects 
are brown dwarfs according to the models by \citet{Baraffe98} (5~Myr) 
which pass all our membership criteria and thus the ratio of substellar
 objects bearing disks increases to 50~\% (note that we need detections 
in at least three IRAC bands to calculate the IRAC slope).

None of our brown dwarf candidates have been detected at 24 micron.  
This is probably due to the detection limits for this band.

As a summary, of the 170 objects presented in Paper I, 167 are discussed here
 (the other three are spurious detections or the Spitzer photometry is not reliable).
 Excluding the sources classified as non-members,
there are 22 which cannot be classified due to the incompleteness of the IRAC data,
 95 have been classified as diskless, another two have transition disks, 25 thin disks
and 20 thick disks. All this information has been listed in Table 4. 
Note that
there are nine
objects classified as Class III from color-color diagrams but which have thin
disks according to the SED slopes, and another one 
(LOri156, a very low mass brown dwarf candidate with a very intense H$\alpha$) which has a 
 thick disk based on the slope of the IRAC data.

\subsection{The Spatial distribution of the members}

We have plotted the spatial distribution of our good candidate members
in Figure~11.  Four-point stars represent B stars and $\lambda$ Orionis
(O8III).   The Class~III members (crosses) are approximately randomly
distributed across the survey region.  Both the Class~II sources and
the B stars give the impression of being concentrated into linear
grouping - with most of the B stars being aligned vertically near
RA = 83.8, and a large number of the Class~II sources being aligned
in the East-Southeast direction (plus some less well-organized
alignments running more or less north-south).   It is possible the
spatial distributions are reflective of the birth processes in C69 -
with the youngest objects (the Class~II sources and B stars) tracing
the (former?) presence of dense molecular gas, whereas the Class~III
sources have had time to mix dynamically and they are no longer near the
locations where they were born.


Figure 12 shows three different views of the central portion of the
Spitzer mosaic at 3.6
microns for the C69 region.  Figures 12a and 12b (with 12b being a
blow-up of the center of 12a) emphasize the distribution of Class~II
sources relative to the cluster center; Figure 12c shows the
distribution of our brown dwarf candidates.  The star $\lambda$ Orionis
is at the center of each of these figures.  The object located
south of the star $\lambda$ Orionis is BD+09 879 C (or HD36861 C, a F8
V star), with an angular distance of about 30 arcsec from the close
binary $\lambda$ Orionis AB (the projected distance, if BD+09 879 C is
a cluster member, would be 12,000 AU from the AB pair).  The apparent 
relative lack of cluster candidate members within about 75 arcsec from the 
star $\lambda$ Orionis may be illusory, as this region was ``burned out"
in the optical images of the CFHT1999 survey and is also adversely
affected in our IRAC images.  There are a number of Class~II sources
at about 75-90 arcsec from $\lambda$\ Ori, corresponding to a projected
separation of order 30,000 AU, so at least at that distance
circumstellar disks can survive despite the presence of a nearby
O star.

Regarding the distribution of brown dwarfs, a significant number of
them (30 \%) are within the the inner circle with a diameter of 9
arcmin (our original optical survey covered an area of 42$\times$28 arcmin).
 However, there are substellar members at any distance from the
star $\lambda$ Ori (Figure ~12c), and there is no substantial evidence that the
cluster brown dwarfs tend to be close to the massive central star.

We have estimated the correlation in spatial distribution of
different sets of data: Class II vs. Class III candidates,
objects with any kind of disk (thin, thick and transition)
 vs. diskless objects, and stellar vs. substellar objects
(following the substellar frontier given in Table 5 for
 different ages and bands).
We have computed the two-sided Kuiper statistic (invariant
Kolmogorov-Smirnov test), and its associated probability that
any of the previously mentioned pairs of stellar groups were drawn
from the same distribution. We have calculated the two
dimensional density function of each sample considering a
 4.5$\times$3 arcmin grid-binning in a 45$\times$30 arcmin region centered at 05:35:08.31,
+09:56:03.6 (the star Lambda Orionis).
The test reveals that in the first case, the cumulative
 distribution function of Class II candidates
is significantly different from that of Class III candidates, with a
 probability for these data sets being drawn from the same
distribution of $\sim 1 \%$. This situation changes when
comparing the set of objects harbouring any kind
of disk with that of diskless objects, finding a probability
 of $\sim 50 \%$ in this case (and hence no conclusion can be
drawn, other than that there is no strong correlation).
On the other hand, regarding a correlation with age, the test points out a
trend in the relationship between the spatial distribution
of stellar and substellar objects depending on the assumed age.
The value of the probability
of these two populations sharing a common spatial distribution
decreases from a  $\sim 30 \%$ when assuming an age of 3 Myr, to $\sim 0.001 \%$
for an age of 8 Myr. The value assuming an age of 5 Myr is $\sim 1 \%$.

The spatial distribution of objects detected at 24 micron can be seen
in Figure~13.  The nebulosity immediately south of the star $\lambda$ Orionis
(close to BD $+$09 879 C) corresponds to the HII region LBN
194.69-12.42 (see the detail in Figure 12b in the band [3.6]).  Most of
the detected members are located within the inner 9 arcmin circle,
with an apparent concentration in a ``filament'' running approximately
north-south (i.e. aligned with the B stars as illustrated in Figure 11b).
Out of the cluster members discovered by
\citeauthor{Dolan01}, 11 are within the MIPS [24] image (see 
Figure~13) and have fluxes above the detection level.  The closest
member to $\lambda$ Orionis is D\&M\#33 (LOri034), about 2 arcmin
east from the central star.

The MIPS image at 24 micron suggest that there are two bubbles
centered around the $\lambda$ Orionis multiple star 
(actually, the center might be the C component or the HII region 
LBN 194.69-12.42). The first one is
about 25 arcmin away, and it is located along the
North-East/South-West axis. More conspicuous is the smaller front
located at a distance of 10.75 arcmin, again centered on the HII
region and not in $\lambda$ Orionis AB.  In this case, it is most visible
located in the direction West/North-West, opposite to the alignment of Class
II objects and low mass members with excess at 24 micron. Similar structures 
can be found at larger scales in the IRAS images of this region,
 at 110 and 190 arcmin. 

 The star 37 Ori, a B0III,  is located at the center of the cocoon at 
the bottom of the image. 
The source  IRAS 05320+0927 is very close and it is  probably the same.

Note that while BD $+$09 879 C would appear to be the source of a
strong stellar wind and/or large UV photon flux, it is not obvious that
the visible star is the UV emitter because the spectral type for
BD $+$09 879 C is given as F8V \citep{Lindroos85}.  It would be useful
to examine this star more closely in order to try to resolve this
mystery.


\section{Conclusions}

We have obtained Spitzer IRAC and MIPS data of an area about one
sq.deg around the star $\lambda$ Orionis, the central star of the 5
Myr Lambda Orionis open cluster (Collinder 69). These data were
combined with our previous optical and near infrared photometry (from
2MASS). In addition, we have obtained deep near infrared imaging. The
samples have been used to assess the membership of the 170 candidate
members, selected from \citet{Barrado04}.

By using the Spitzer/IRAC data and the criteria developed by Allen et
al. (2004) and Hartmann et al. (2005), we have found 33 objects which
can be classified as Classical TTauri stars and substellar analogs
(Class~II objects). This means that the fraction of members with disks
is 25\% and 40\%, for the stellar (in the spectral range M0 - M6.5) 
and substellar population 
(down to 0.04 M$_\odot$). However,
by combining this information with H$\alpha$\ emission (only a
fraction of them have spectroscopy), we find that some do not seem
to be accreting.

Moreover, as expected from models,
  we see a correlation in the $[3.6]$ - $[4.5]$ vs. $[5.8]$ - $[8.0]$ diagram
  for objects with redder colors (more IR excess) to have stronger
  H$\alpha$\  emission.
In addition, following \citet{Lada06} and the classification 
based on the slope of the IRAC data, we found that the ratio of 
substellar members 
bearing disks (optically thin or thick) is $\sim$~50\%, whereas is 
about 31\% for the complete sample (14\% with thick disks). 
This result suggests that the timescale for primordial disks 
to dissipate is longer for lower
  mass stars, as suggested in Barrado y Navascu\'es  \& Mart\'{\i}n (2003).

We have also found that the distribution of Collinder members is very
inhomogeneous, specifically for the Class~II objects. Most of them are
located in a filament which goes from the central star $\lambda$
Orionis to the south-east, more or less towards the dark cloud Barnard
35. In addition, there are several Class~II stars close to the central
stars.   If the (previously) highest mass member of C69 has already
evolved off the main sequence and become a supernova, either the
disks of these Class~II stars survived that episode or they formed
subsequent to the supernova.

We have also derived the fluxes at 24 micron from Spitzer/MIPS
imaging. Only a handful --13-- of the low mass stars were detected (no
brown dwarfs). Most of them are Class~II objects. In the case of the
two Class~III members with 24 micron excess, it seems that they
correspond to transitions disks, already evolving toward the
protoplanetary phase.

\acknowledgements
 We thank Calar Alto Observatory for allocation of director's discretionary 
time to this programme. 
This research has been funded by Spanish grants  MEC/ESP2004-01049, MEC/Consolider-CSD2006-0070, 
and CAM/PRICIT-S-0505/ESP/0361.



%
%
\clearpage

\begin{deluxetable}{lllll}
\setlength{\tabcolsep}{0.02in}
\tabletypesize{\tiny}
\tablewidth{0pt}
\tablecaption{Additional near infrared photometry for the candidate
 members of the Lambda Orionis cluster (WHT/INGRID).}
\tablehead{
\colhead{ Name}   & 
\colhead{ I   error}  &  
\colhead{ J   error}   & 
\colhead{ H   error}    & 
\colhead{Ks   error}   
}
\startdata
 LOri006 & 12.752 & 11.67  0.01 & 10.90  0.01 & 10.94  0.01 \\    
 LOri007 & 12.779 & 11.65  0.01 & --     --   & --     --   \\    
 LOri008 & 12.789 & 11.53  0.01 & 10.85  0.01 & 10.62  0.01 \\    
 LOri009 & 12.953 & 11.79  0.01 & --     --   & --     --   \\    
 LOri011 & 13.006 & 11.59  0.01 & --     --   & --     --   \\    
 LOri015 & 13.045 & 11.90  0.01 & --     --   & --     --   \\    
 LOri016 & 13.181 & 12.00  0.01 & 11.41  0.01 & 11.27  0.01 \\    
 LOri020 & 13.313 & 11.95  0.01 & --     --   & --     --   \\
\enddata
$\,$ \\

\end{deluxetable}

\begin{deluxetable}{lcccc}
\setlength{\tabcolsep}{0.02in}
\tabletypesize{\scriptsize}
\tablewidth{0pt}
\tablecaption{Zero points,  aperture corrections and conversion factors between the magnitudes and the
fluxes in  Jansky.}
\tablehead{
\colhead{Channel}         & 
\colhead{Ap. correction ap=4px} (mag) & 
\colhead{Ap. correction ap=2px} (mag) & 
\colhead{Zero Point (mag)\tablenotemark{a}}     & 
\colhead{Flux mag=0 (Jy)} \\
}
\startdata
$[3.6]$ &  0.090 &  0.210 & 17.26 &  280.9\tablenotemark{b} \\
$[4.5]$ &  0.102 &  0.228 & 16.78 &  179.7\tablenotemark{b} \\
$[5.8]$ &  0.101 &  0.349 & 16.29 &  115.0\tablenotemark{b} \\
$[8.0]$ &  0.121 &  0.499 & 15.62 &   64.1\tablenotemark{b} \\
$[24]$  &  \multicolumn{2}{c}{0.168\tablenotemark{d}} & 11.76 &    7.14\tablenotemark{c}\\
\enddata
\tablenotetext{a}{Zero Points for aperture photometry performed with IRAF on
  the BCD data.}
\tablenotetext{b}{\citet{Reach05}}
\tablenotetext{c}{\citet{Engelbracht06}}
\tablenotetext{d}{The aperture used for MIPS [24] was always 5.31 pixels.}
\end{deluxetable}

\begin{deluxetable}{llllllllllll}
\setlength{\tabcolsep}{0.02in}
\tabletypesize{\scriptsize}
\rotate
\tablewidth{0pt}
\tablecaption{Candidate members of the Lambda Orionis cluster (Collinder 69)}
\tablehead{
\colhead{ Name}   & 
\colhead{ R     error}   & 
\colhead{ I     error}   &  
\colhead{ J     error}   & 
\colhead{ H     error}   & 
\colhead{Ks     error}   & 
\colhead{[3.6]  error}   & 
\colhead{[4.5]  error}   & 
\colhead{[5.8]  error}   & 
\colhead{[8.0]  error}   & 
\colhead{[24]   error}   & 
\colhead{Mem\tablenotemark{1}} \\ 
}
\startdata
LOri001 & 13.21 0.00 & 12.52 0.00 & 11.297 0.022 & 10.595 0.022 & 10.426 0.021 & 10.228 0.003     & 10.255 0.004     & 10.214 0.009     & 10.206 0.010     & --     --     & Y      \\ 
LOri002 & 13.44 0.00 & 12.64 0.00 & 11.230 0.024 & 10.329 0.023 & 10.088 0.019 &  9.935 0.003     & 10.042 0.003     &  9.930 0.009     &  9.880 0.008     & --     --     & Y        \\ 
LOri003 & 13.39 0.00 & 12.65 0.00 & 11.416 0.023 & 10.725 0.022 & 10.524 0.023 & 10.262 0.003     & 10.318 0.004     & 10.239 0.010     & 10.171 0.010     & --     --     & Y      \\ 
LOri004 & 13.71 0.00 & 12.65 0.00 & 11.359 0.022 & 10.780 0.023 & 10.548 0.021 & 10.287 0.003     & 10.249 0.004     & 10.185 0.009     & 10.127 0.009     & --     --     & Y        \\ 
LOri005 & 13.38 0.00 & 12.67 0.00 & 11.378 0.022 & 10.549 0.022 & 10.354 0.023 & 10.204 0.003     & 10.321 0.004     & 10.218 0.009     & 10.158 0.009     & --     --     & Y        \\ 
LOri006 & 13.55 0.00 & 12.75 0.00 & 11.542 0.026 & 10.859 0.026 & 10.648 0.021 & 10.454 0.003     & 10.454 0.004     & 10.399 0.011     & 10.319 0.010     & --     --     & Y       \\ 
LOri007 & 13.72 0.00 & 12.78 0.00 & 11.698 0.027 & 11.101 0.024 & 10.895 0.030 & 10.668 0.004     & 10.636 0.004     & 10.615 0.012     & 10.482 0.013     & --     --     & Y        \\ 
LOri008 & 13.60 0.00 & 12.79 0.00 & 11.548 0.029 & 10.859 0.023 & 10.651 0.024 & 10.498 0.003     & 10.495 0.004     & 10.440 0.011     & 10.256 0.012     & --     --     & Y      \\
\enddata 
\tablenotetext{1}{Final membership.}
\tablenotetext{2} {2-pixel aperture radius used for the photometry due to the presence of nearby objects or hot pixels.}
\tablenotetext{*}{LOri097 and LOri145 are artifacts. LOri123 is a non member and has uncertainties in its photometry.}\\
\end{deluxetable}                                                              


\begin{deluxetable}{lllllllll}
\setlength{\tabcolsep}{0.02in}
\tabletypesize{\scriptsize}
\tablewidth{0pt}
\tablecaption{Candidate members of the Lambda Orionis cluster (Collinder 69)}
\tablehead{
\colhead{ Name}   & 
\colhead{ SpT}           &
\colhead{Phot.Mem\tablenotemark{1}}  &
\colhead{Mem\tablenotemark{2}}      & 
\colhead{Mem\tablenotemark{3}}       & 
\colhead{IRAC classification\tablenotemark{4}}  &
\colhead{SED slope\tablenotemark{5}}  &
\colhead{Disk type}  &
\colhead{Comment\tablenotemark{6}} \\
}
\startdata
LOri001 & --    & Y  Y  Y  Y  Y  Y  Y  & Mem  & Y  & III                 & -2.8  & Diskless   & Ha-   WHa(DM)=2.51  DM\#01        \\ 
LOri002 & --    & Y  Y  Y  Y  Y  Y  Y  & NM-  & Y  & III                 & -2.74 & Diskless   & --                               \\ 
LOri003 & --    & Y  Y  Y  Y  Y  Y  Y  & Mem  & Y  & III                 & -2.71 & Diskless   & Ha-   WHa(DM)=3.35  DM\#46        \\ 
LOri004 & --    & Y  Y  Y  Y  Y  Y  Y  & Mem  & Y  & III                 & -2.65 & Diskless   & --                               \\ 
LOri005 & --    & Y  Y  Y  Y  Y  Y  Y  & NM-  & Y  & III                 & -2.75 & Diskless   & --                               \\ 
LOri006 & --    & Y  Y  Y  Y  Y  Y  Y  & Mem  & Y  & III                 & -2.68 & Diskless   & --                               \\ 
LOri007 & --    & Y  Y  Y  Y  Y  Y  Y  & Mem  & Y  & III                 & -2.63 & Diskless   & --                               \\ 
LOri008 & --    & Y  Y  Y  Y  Y  Y  Y  & Mem  & Y  & III                 & -2.56 & Diskless   & Ha-   WHa(DM)=1.65  DM\#51        \\
\enddata 
\tablenotetext{1}{Membership is Ivs(I-J); Ivs(I-K); Ivs(I-3.6); Ivs(I-4.5); Jvs(J-3.6);Kvs(K-3.6); Jvs(J-K).} 
\tablenotetext{2}{Membership as in Paper~I.} 
\tablenotetext{3}{Final membership.}
\tablenotetext{4} {Classification as measured in the IRAC CCD --[3.6]-[4.5]
    versus [5.8]-[8.0]. Class~III stands for diskless members and Class~II are
    Classical TTauri stars or substellar analogs.}
\tablenotetext{5}{IRAC slope. \citet{Lada06} classified the objects according
  to their IRAC slope: $\alpha<$-2.56 for a diskless object,
  -2.56$<\alpha<$-1.80 for a transition object, and $\alpha>$-1.80 for objects
bearing optically thick disks}
\tablenotetext{6}{Ha+ = W(Halpha) above the saturation criterion. \\
Ha- = W(Halpha) below the saturation criterion.  \\   
200km =  width of Halpha equal or larger than this value.\\
WHa(DM) = from Dolan \& Mathieu}
\tablenotetext{7}{{\bf Probably diskless objects. The different results on
    IRAC CCD and IRAC slope are probably due to an uncertain measure at 5.8~$\mu$m.}}
\tablenotetext{8}{{\bf Probably a class II source with an uncertain measure at 8.0~$\mu$m.}}
\end{deluxetable}                                                              

\begin{deluxetable}{ccccc}
\setlength{\tabcolsep}{0.02in}
\tabletypesize{\scriptsize}
\tablewidth{0pt}
\tablecaption{Location of the substellar frontier, using models by
 Baraffe et al. (1998) and a distance of 400 pc. 
Values such as 340 or 450 pc would modify the listed magnitudes by
$-$0.35 and $+$0.26, respectively. We have included an
interstellar reddening of $E(B-V)$=0.12, equivalent to
 $A_I$=0.223, $A_J$=0.106, $A_K$=0.042, $A_L$=0.022
}
\tablehead{
\colhead{Age (Myr)} & 
\colhead{$Ic$}      & 
\colhead{$J$}       & 
\colhead{$Ks$}      & 
\colhead{$L'$}     
}
\startdata
 1 &  16.72 &  14.35 &  13.32 &  12.88  \\
 3 &  17.18 &  14.87 &  13.84 &  13.40  \\
 5 &  17.55 &  15.36 &  14.35 &  13.92  \\
 8 &  17.92 &  15.80 &  14.80 &  14.36  \\
10 &  18.13 &  16.01 &  15.01 &  14.57  \\
16 &  18.52 &  16.40 &  15.40 &  14.96  \\
20 &  18.71 &  16.59 &  15.60 &  15.15  \\
\enddata
\end{deluxetable}


\clearpage

%
%


\setcounter{figure}{0}
    \begin{figure*}
    \centering
    \includegraphics[width=10.8cm]{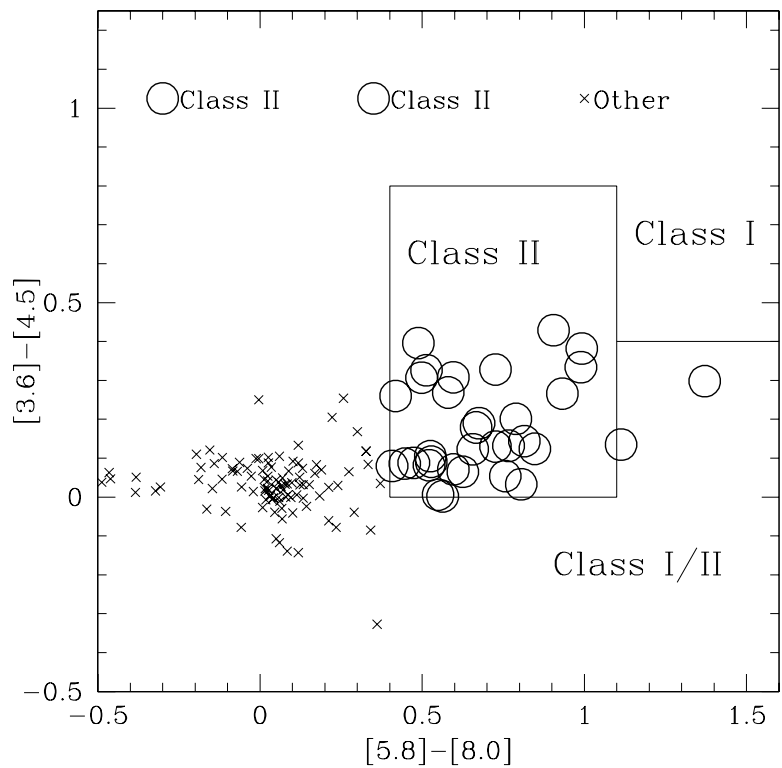}
 \caption{
Spitzer/IRAC CCD. Class~I/II (big empty circles, magenta),
 Class~II (big empty circles, red) and Class~III --or not members-- 
(crosses) have been classified using this diagram 
(After \citet{Allen04} and \citet{Hartmann05}).
}
 \end{figure*}

\setcounter{figure}{1}
    \begin{figure*}
    \centering
    \includegraphics[width=7.8cm]{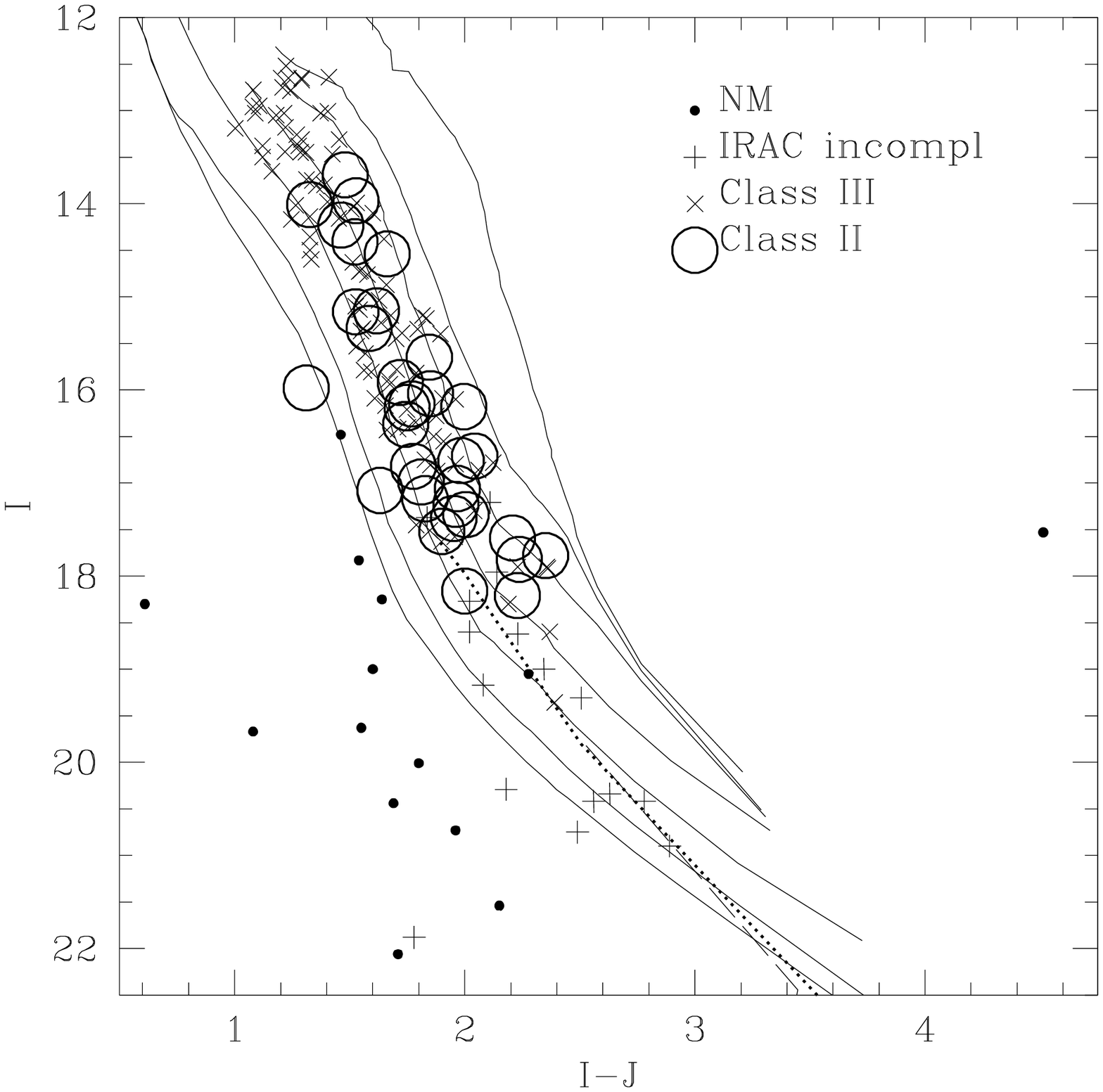}
    \includegraphics[width=7.8cm]{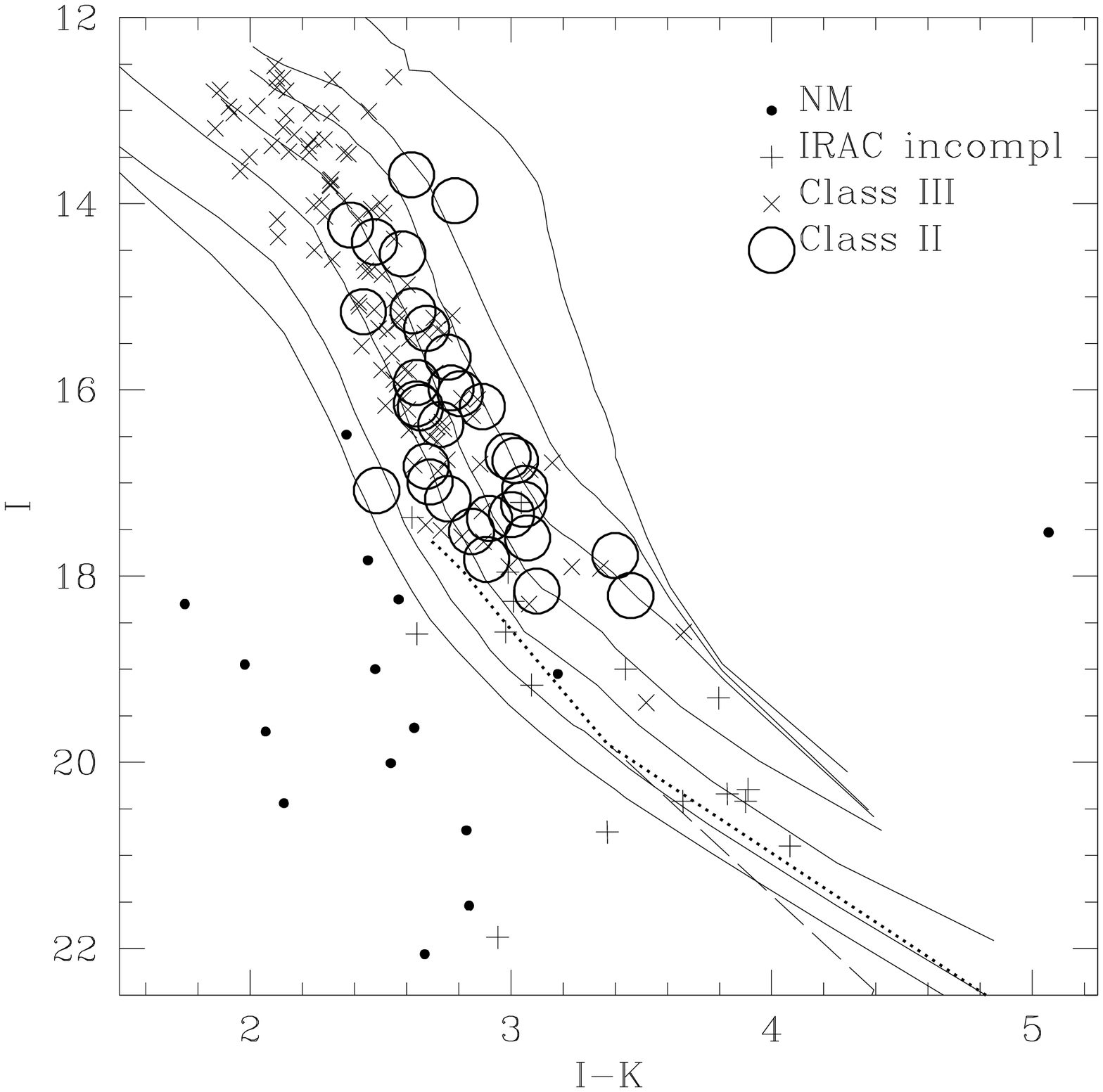}
    \includegraphics[width=7.8cm]{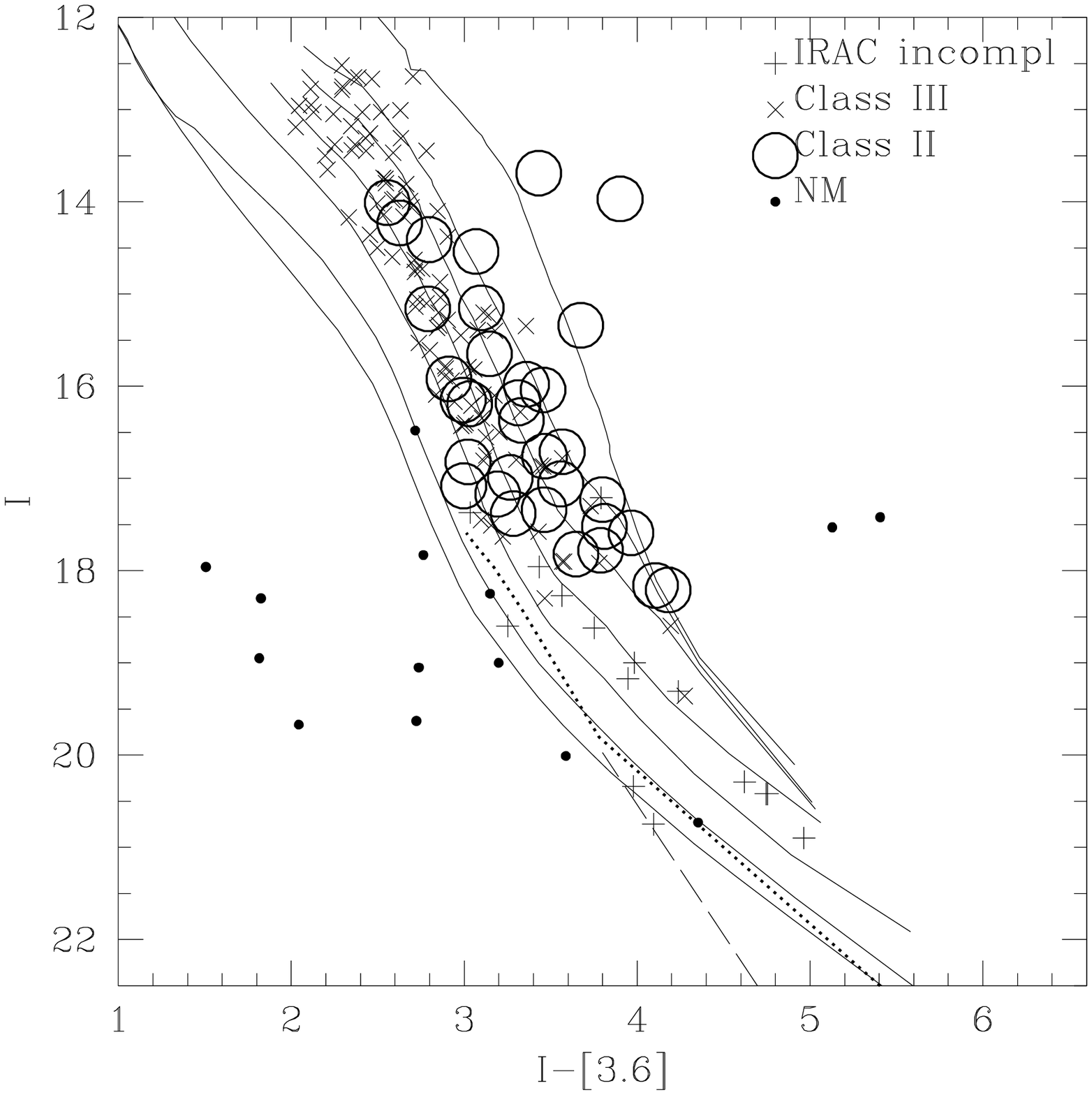}
    \includegraphics[width=7.8cm]{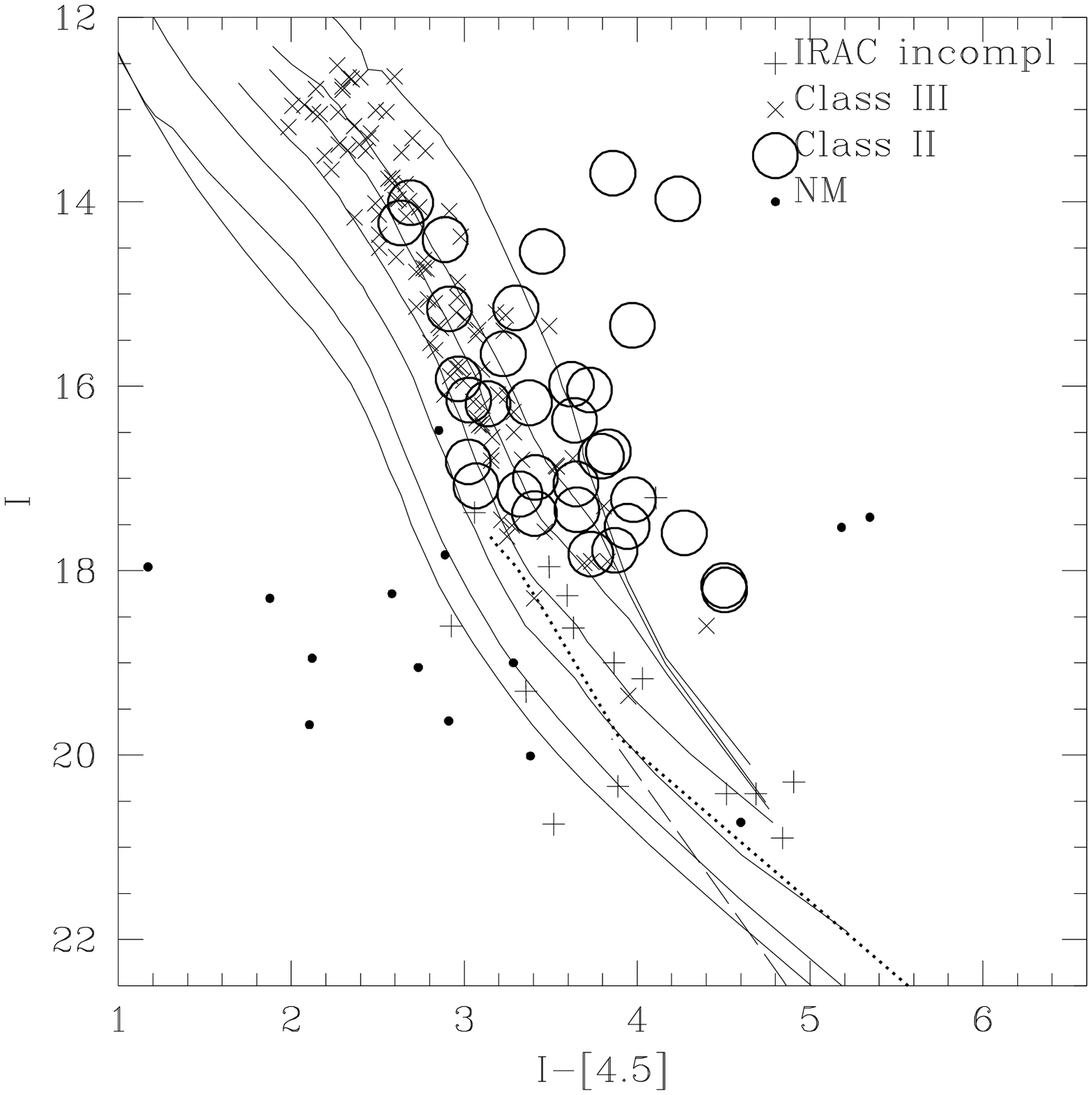}
 \caption{
Optical/IR Color-Magnitude Diagram. Non-members appear as dots. 
Class~II sources 
(Classical TTauri stars and substellar analogs) 
have been included as big (red) circles, whereas 
Class~III (Weak-line TTauri) objects appear as crosses, and other Lambda Orionis
members lacking the  complete set of IRAC photometry are displayed with the  plus
symbol.
The figure includes 1, 3, 5, 10, 20, 50, and 100 Myr isocrones from Baraffe et al. (1998) as 
solid lines, as well as 5 Myr isochrones corresponding to dusty and COND models
\citep{Chabrier00, Baraffe02}, as dotted and dashed lines. 
}
 \end{figure*}


\setcounter{figure}{2}
    \begin{figure*}
    \centering
    \includegraphics[width=7.8cm]{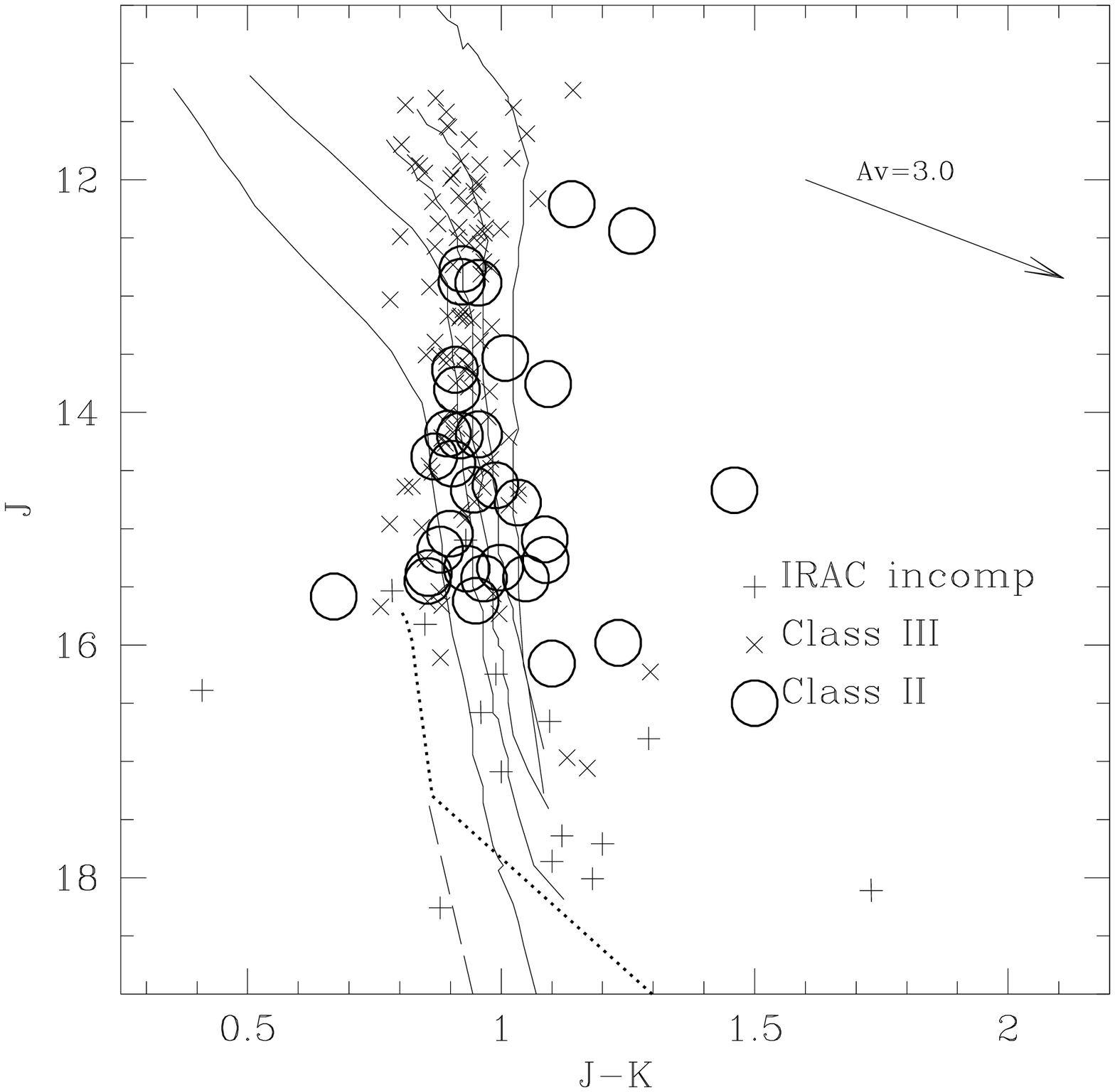}
    \includegraphics[width=7.8cm]{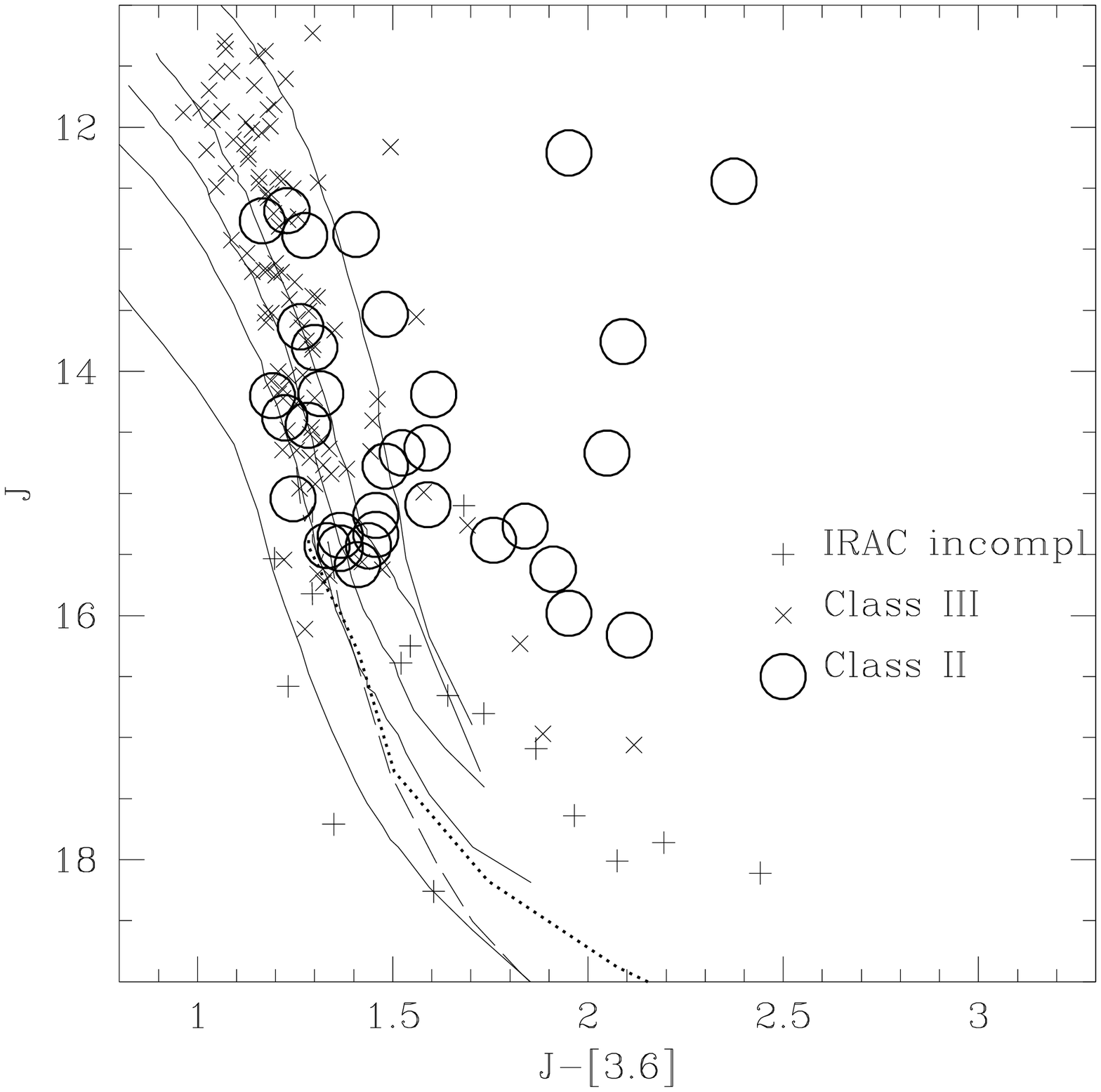}
    \includegraphics[width=7.8cm]{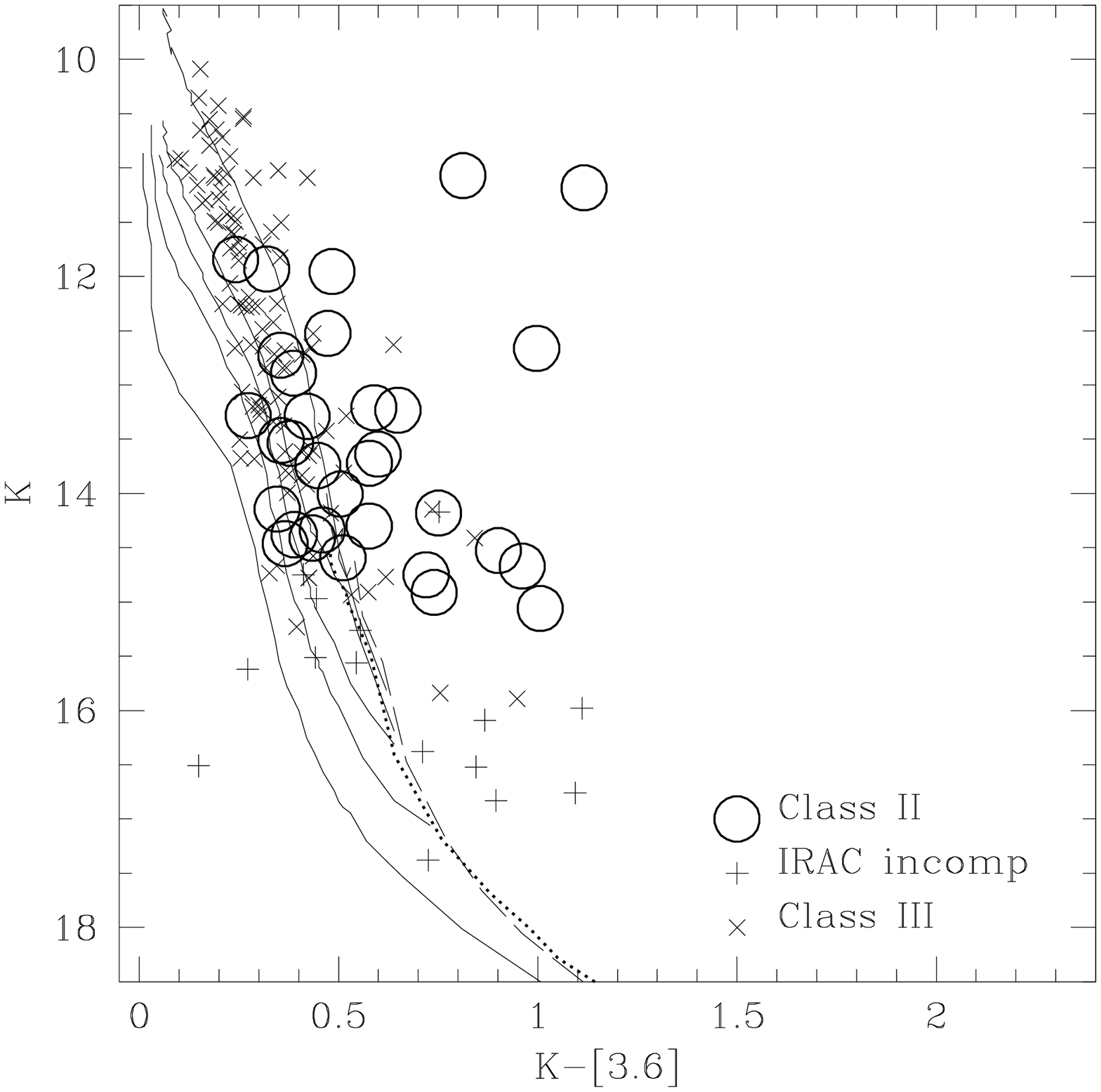}
    \includegraphics[width=7.8cm]{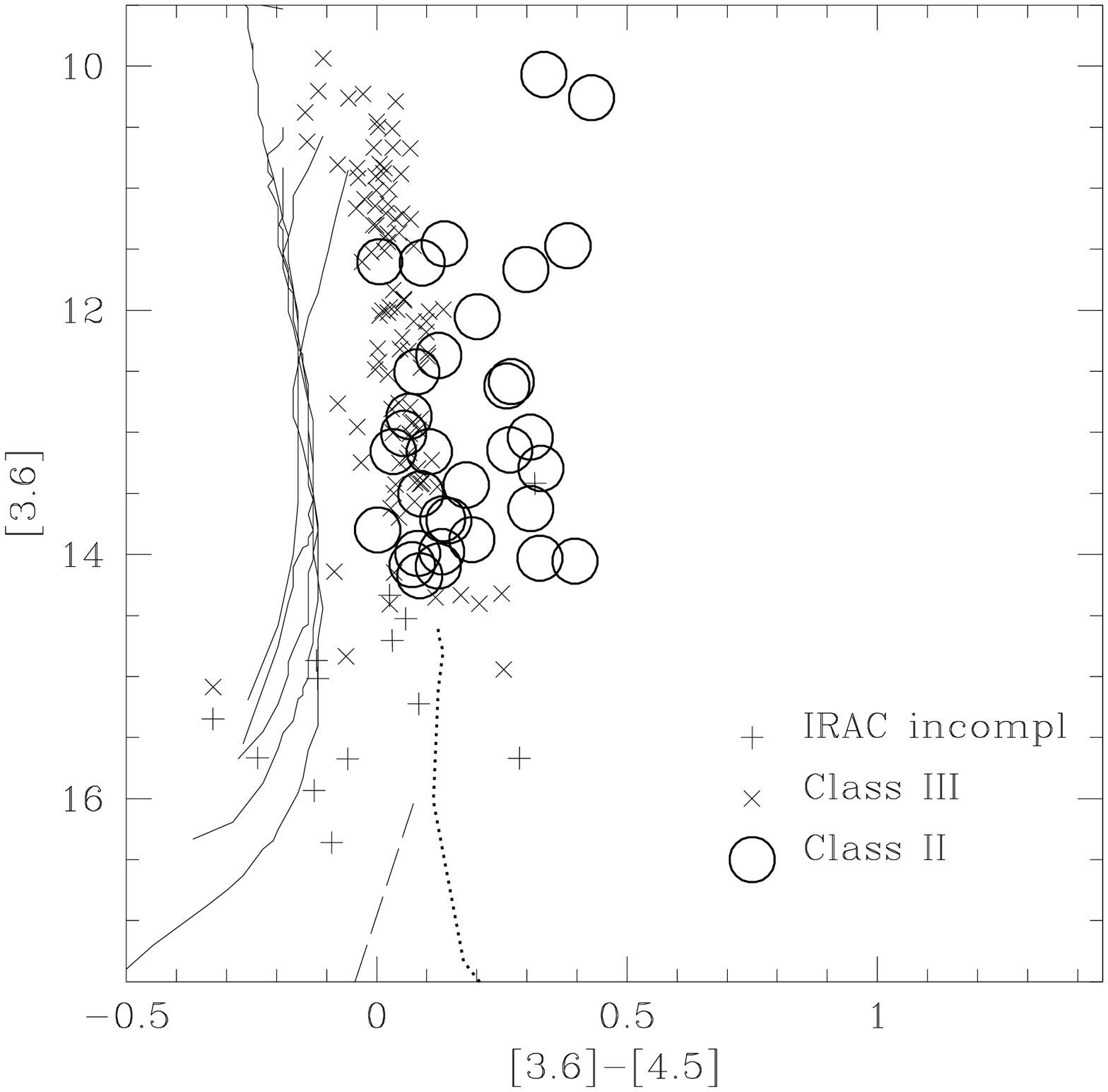}
 \caption{
Near-IR and Spitzer  Color-Magnitude Diagram.  
Class~II sources 
(Classical TTauri stars and substellar analogs) 
have been included as big (red) circles, whereas Class~III 
(Weak-line TTauri) objects appear as crosses, and other Lambda Orionis
members lacking the  complete set of IRAC photometry are displayed with the  plus
symbol.
The figure includes 1, 5, 10, 20, and 100 Myr isochrones from Baraffe et al. (1998)
as 
solid lines, as well as 5 Myr isochrones corresponding to dusty and COND models
(Chabrier et al. 2000; Baraffe et al. 2002), as dotted and dashed lines. 
Note that in the last panel we have the the L and M data
 for the NextGen models, since Spitzer photometry has not been computed for this
set of models.
}
 \end{figure*}


\setcounter{figure}{3}
    \begin{figure*}
    \centering
    \includegraphics[width=16.8cm]{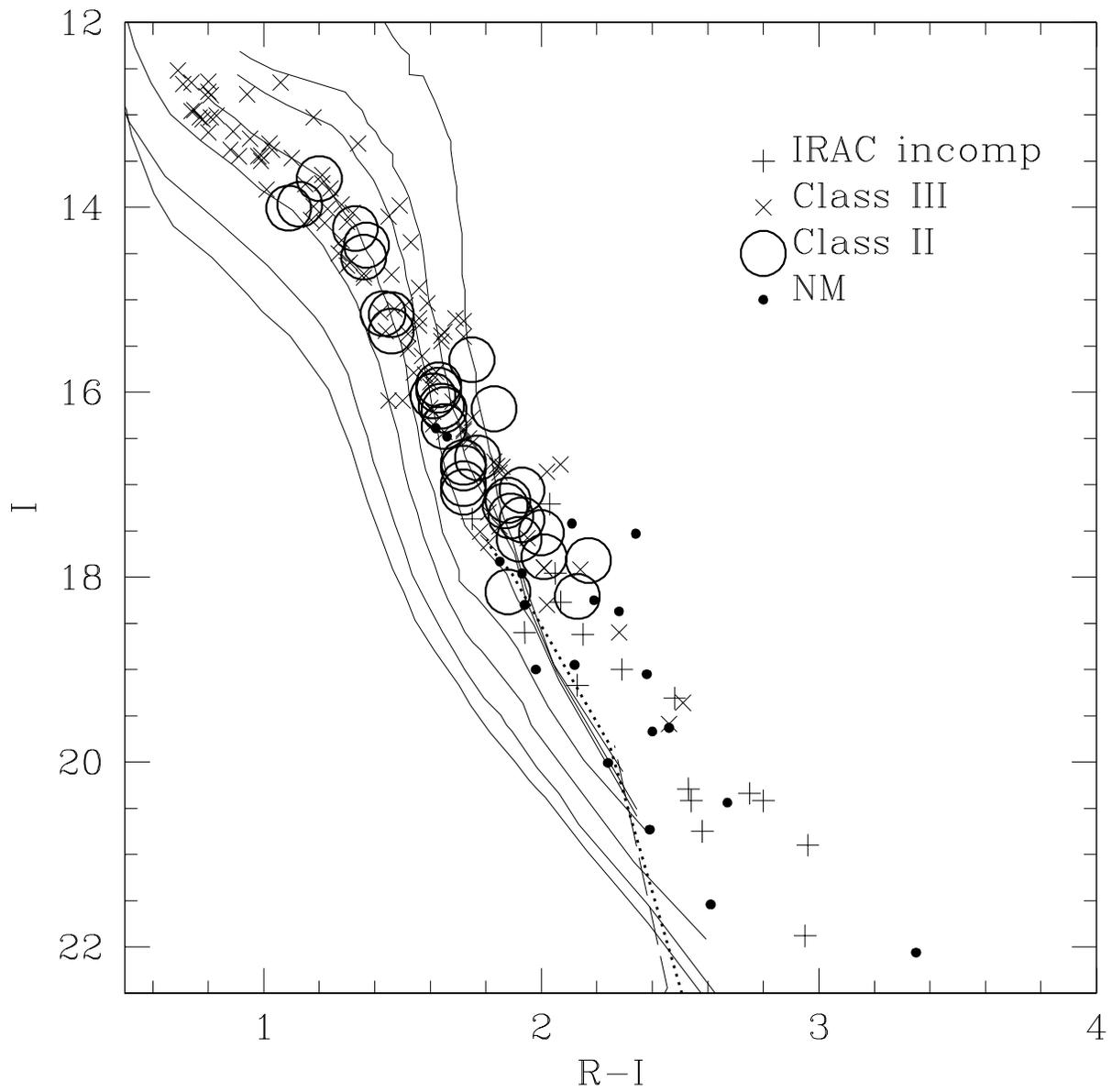}
 \caption{
Optical Color-Magnitude Diagram with the CFHT magnitudes and our new membership 
classification. Symbols as in previous figures. 
}
 \end{figure*}


\setcounter{figure}{4}
    \begin{figure*}
    \centering
    \includegraphics[width=10.8cm]{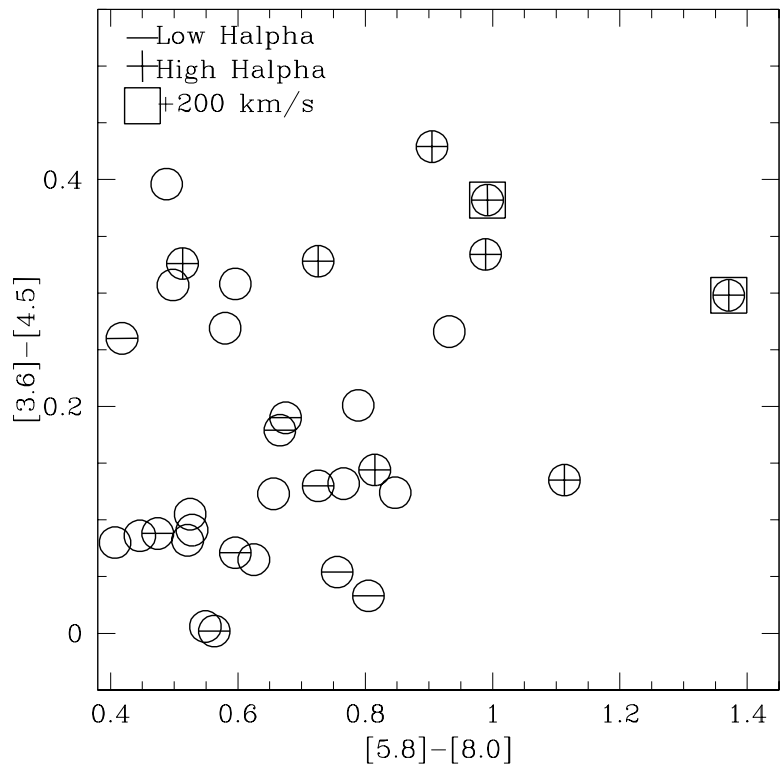}
 \caption{
Spitzer/IRAC CCD for Class~II objects.
We have included information regarding the H$\alpha$\  emission.
}
 \end{figure*}


\setcounter{figure}{5}
    \begin{figure*}
    \centering
    \includegraphics[width=4.2cm]{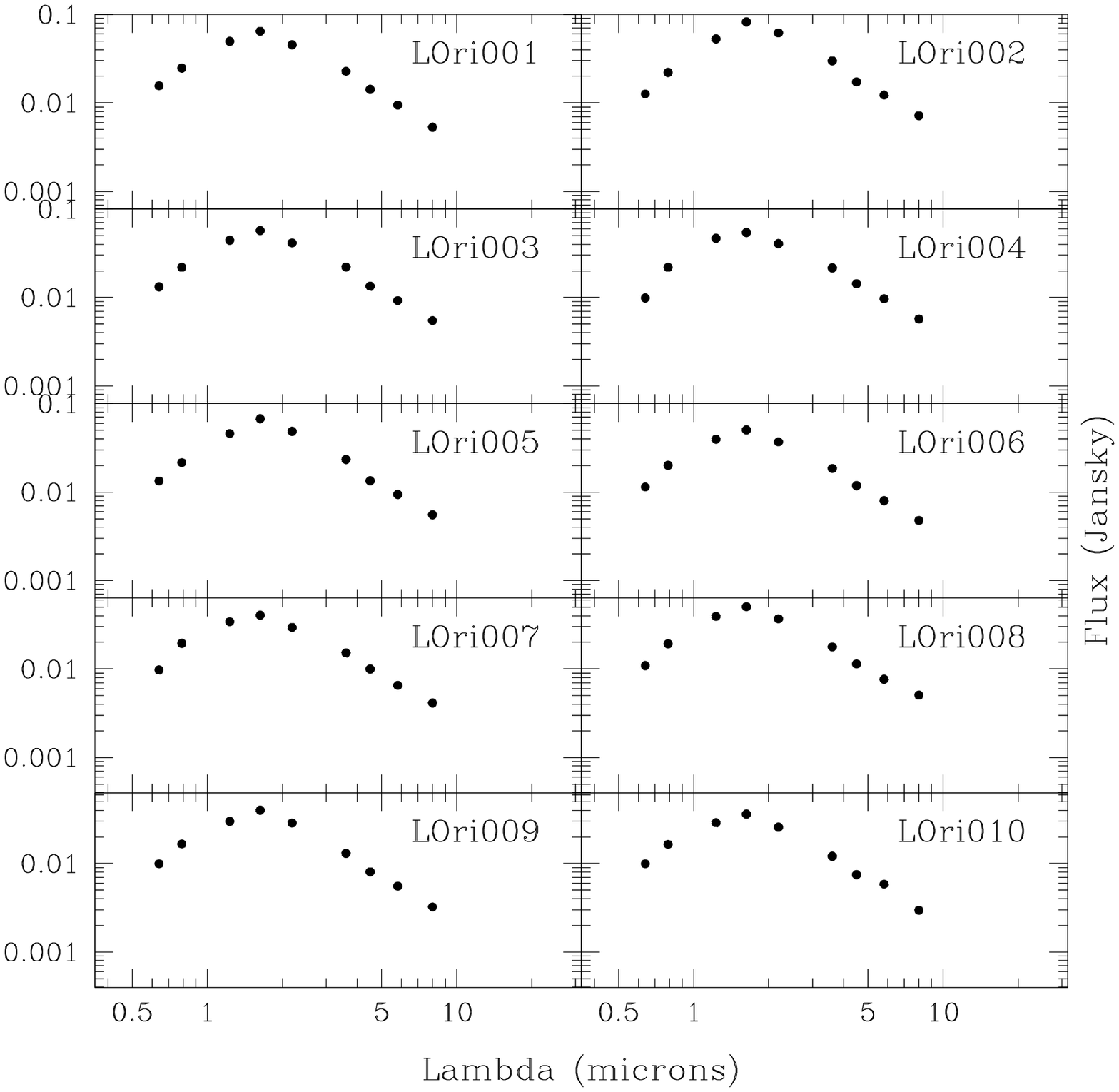}
    \includegraphics[width=4.2cm]{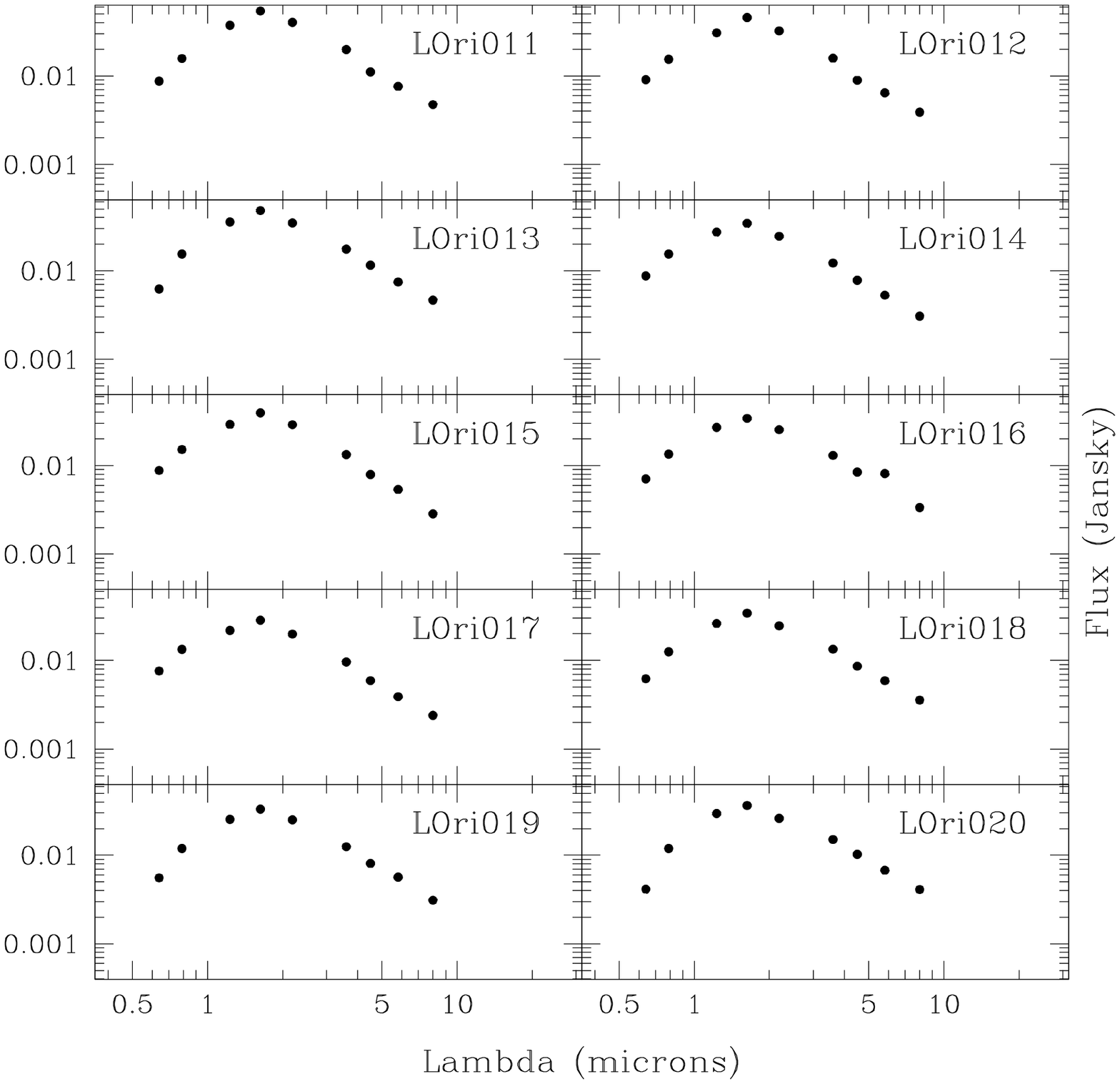}
    \includegraphics[width=4.2cm]{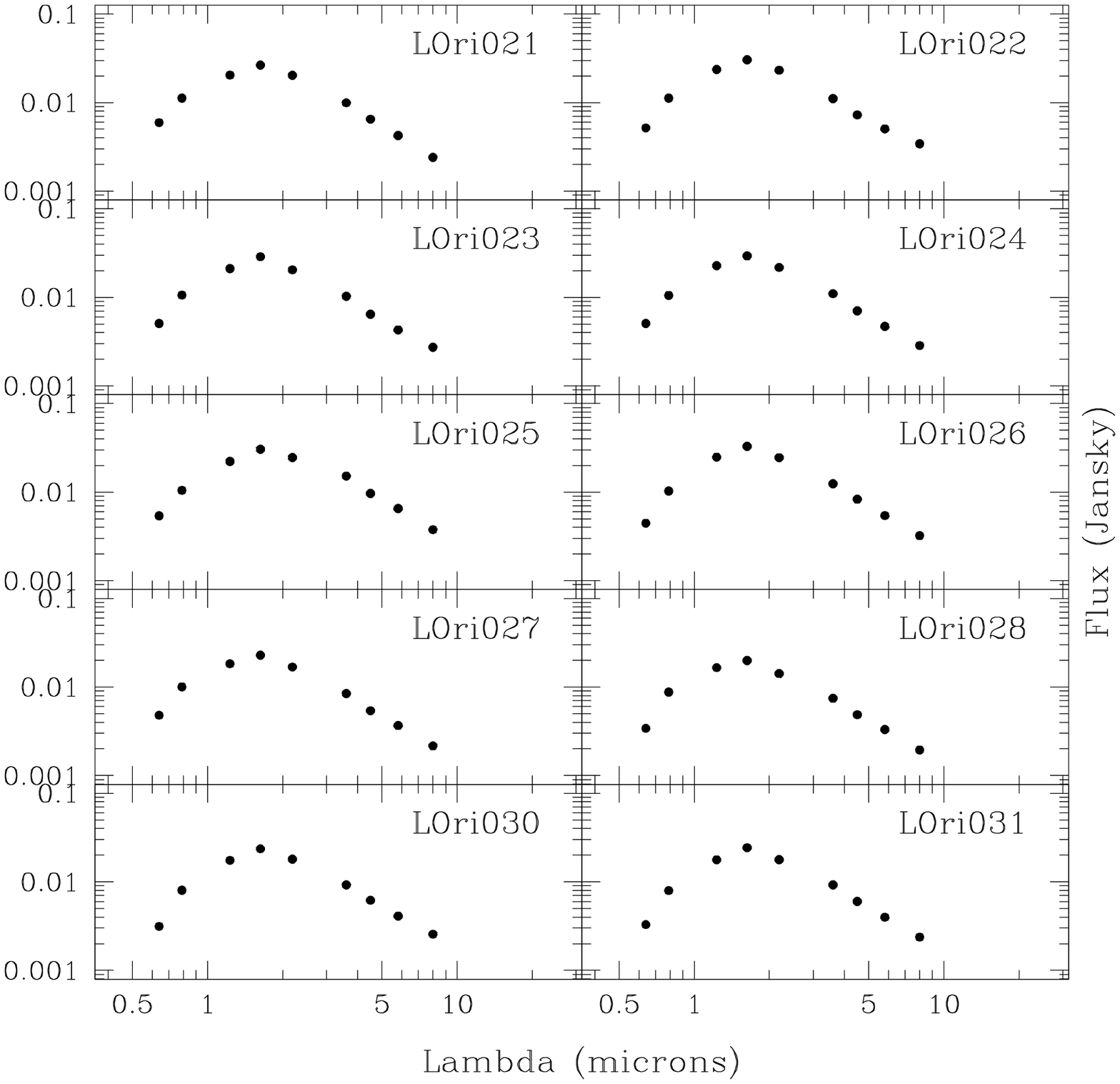}
    \includegraphics[width=4.2cm]{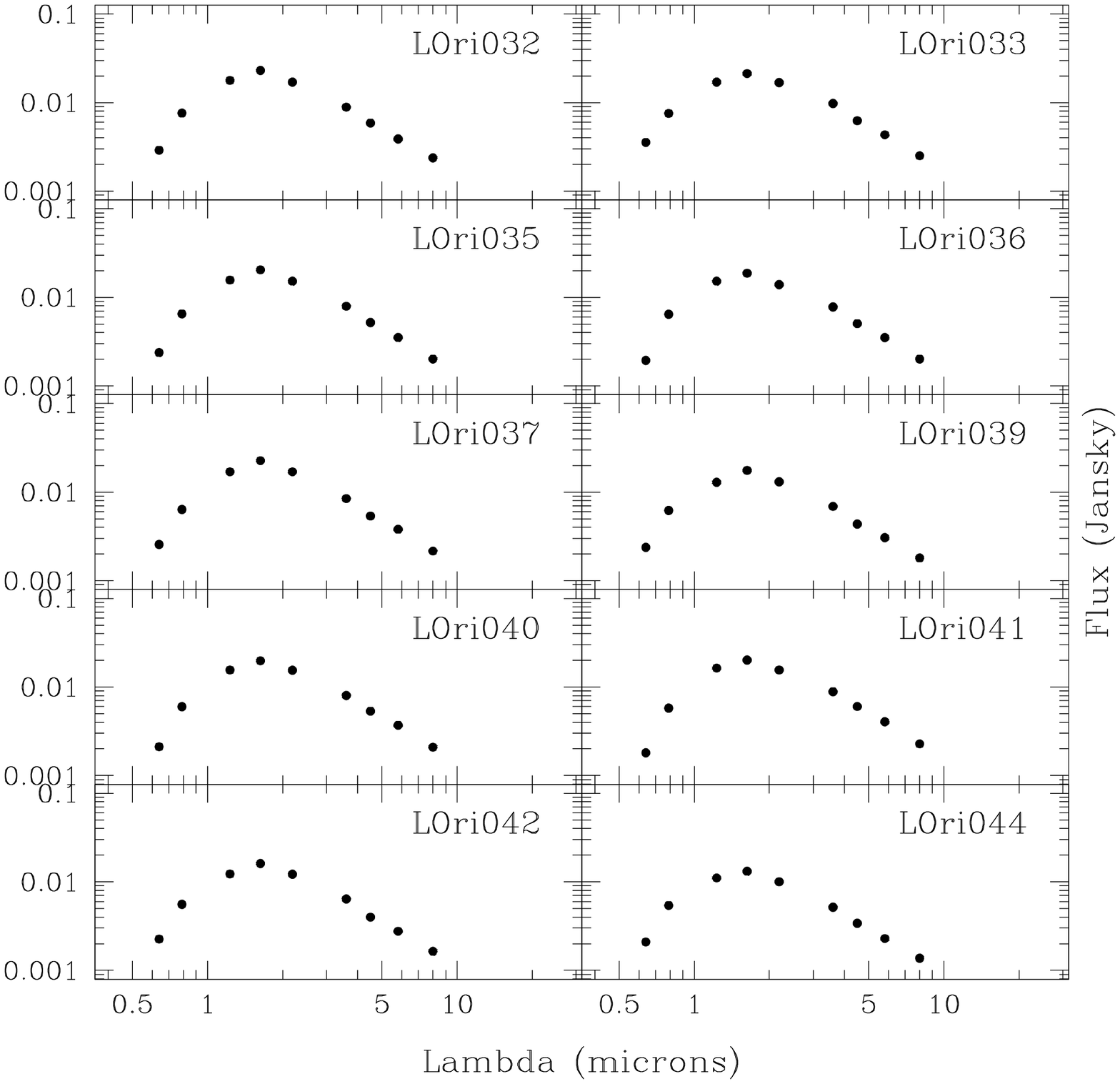}
    \includegraphics[width=4.2cm]{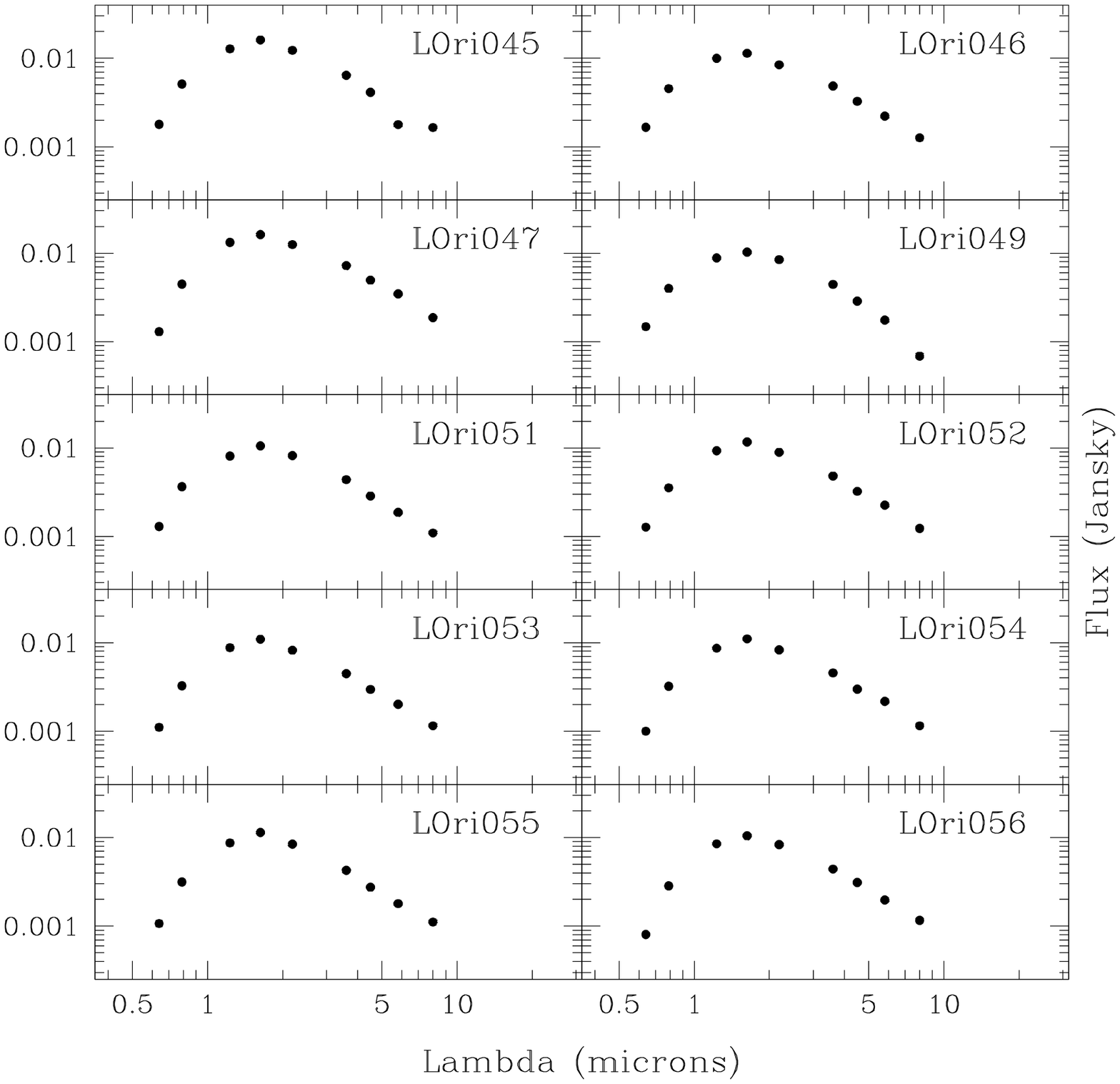}
    \includegraphics[width=4.2cm]{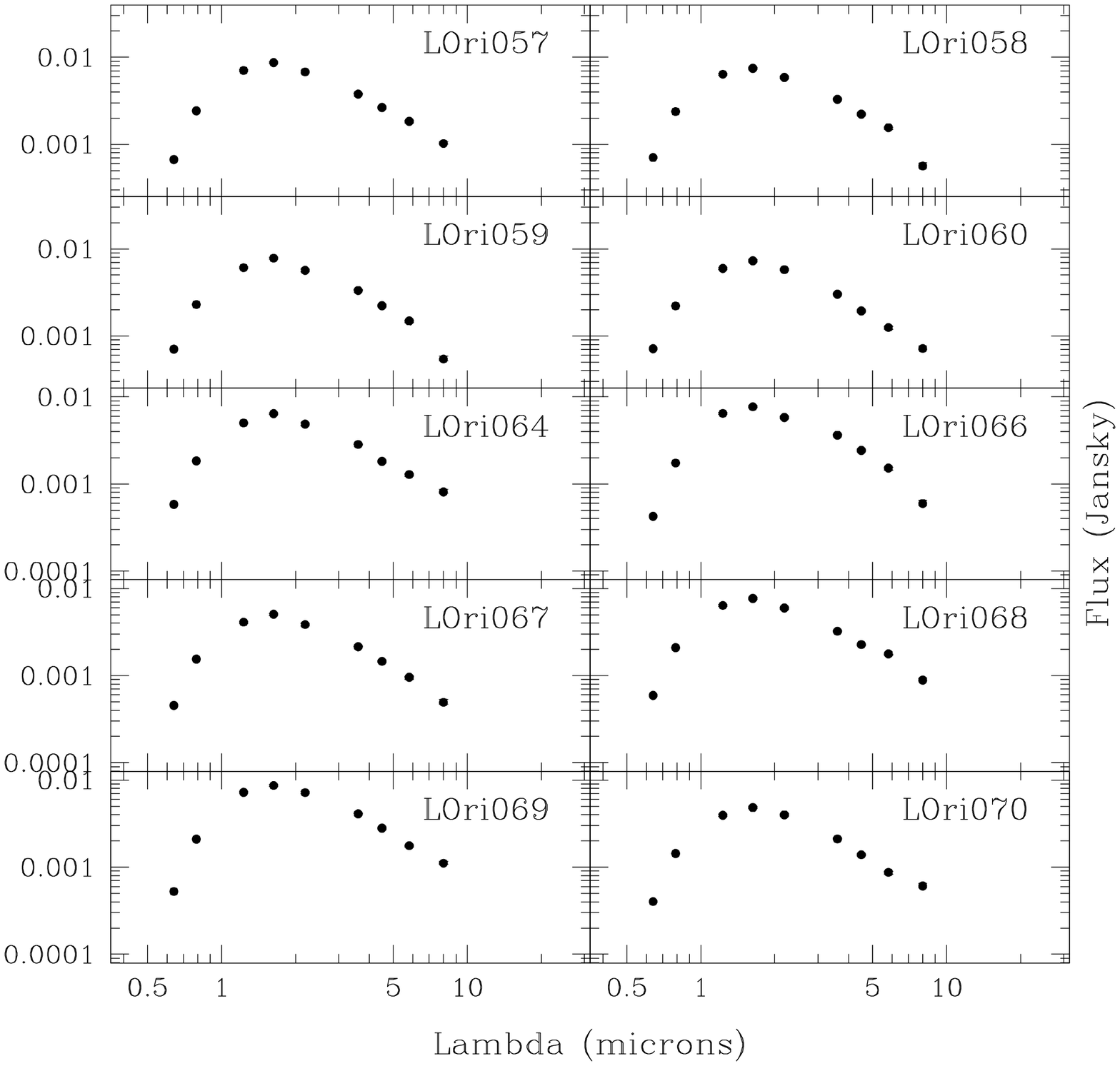}
    \includegraphics[width=4.2cm]{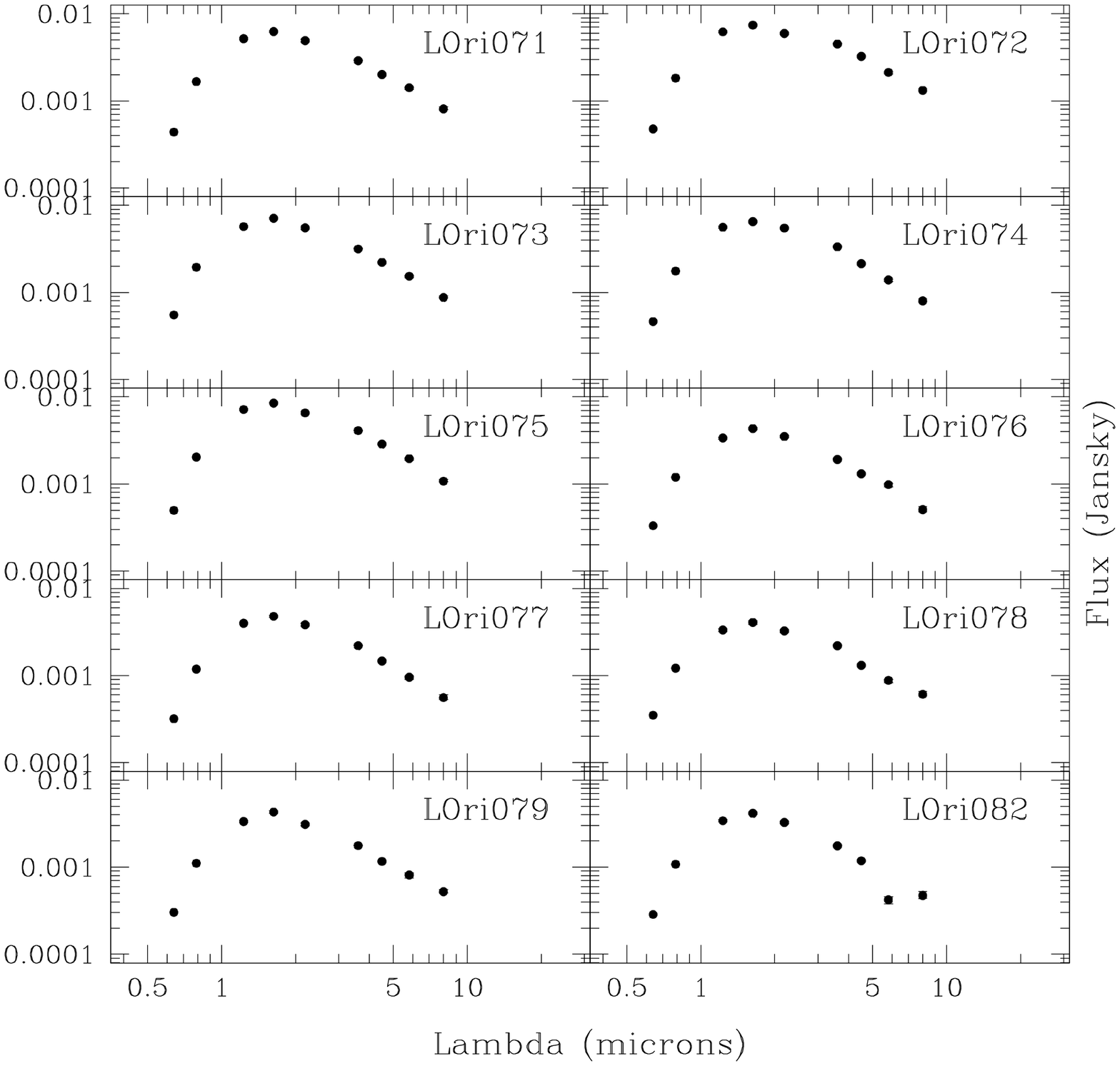}
    \includegraphics[width=4.2cm]{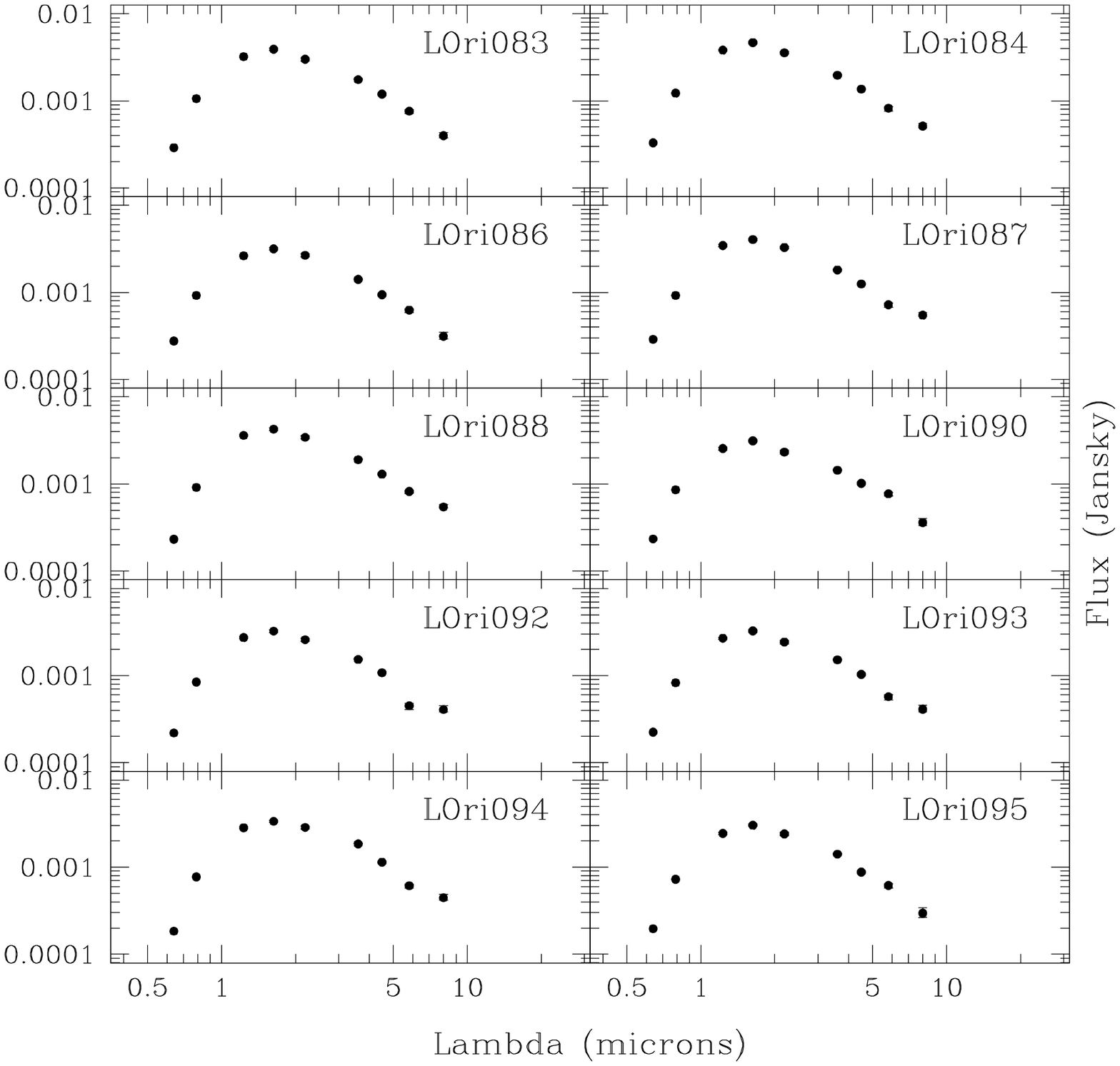}
    \includegraphics[width=4.2cm]{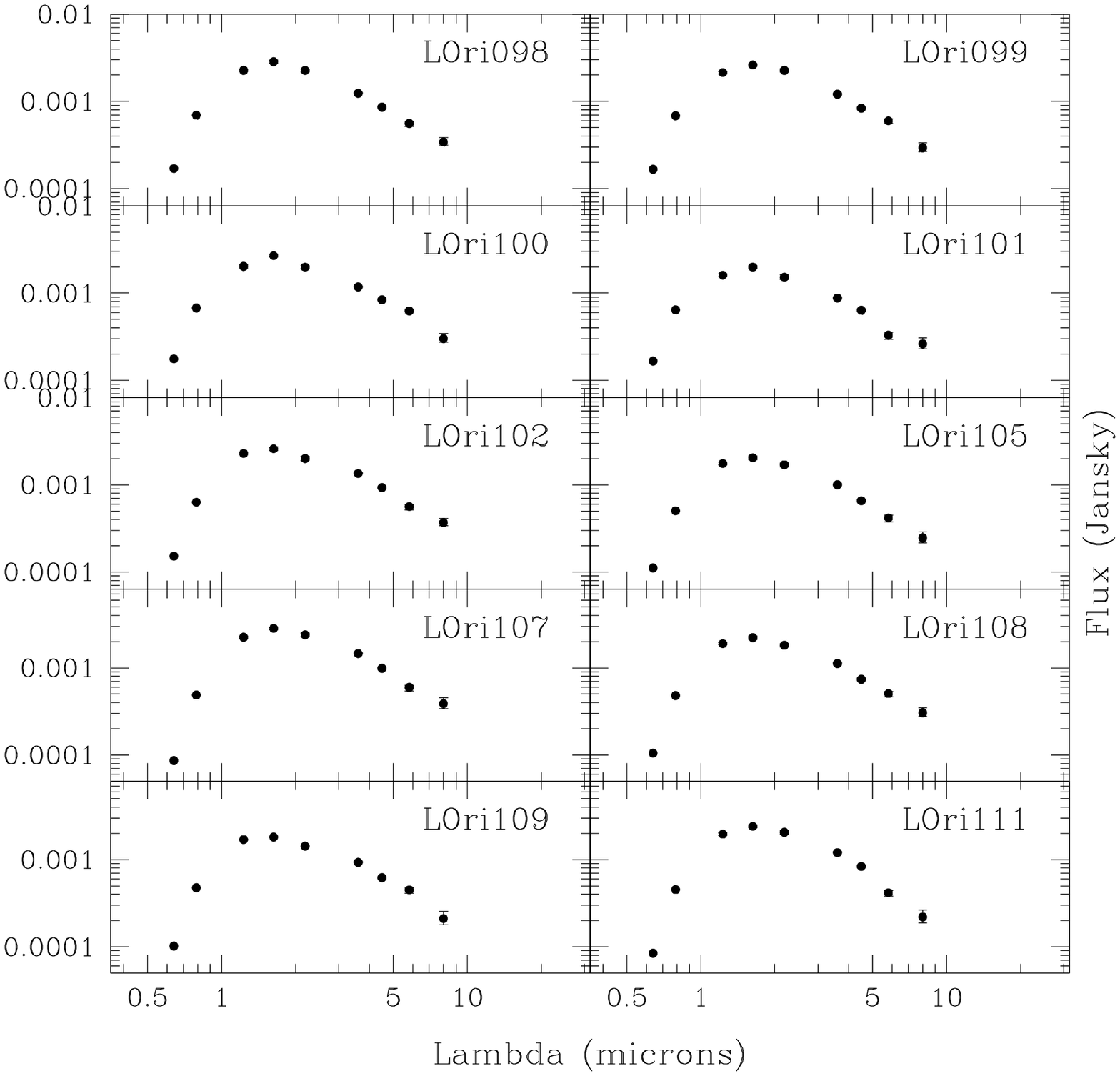}
    \includegraphics[width=4.2cm]{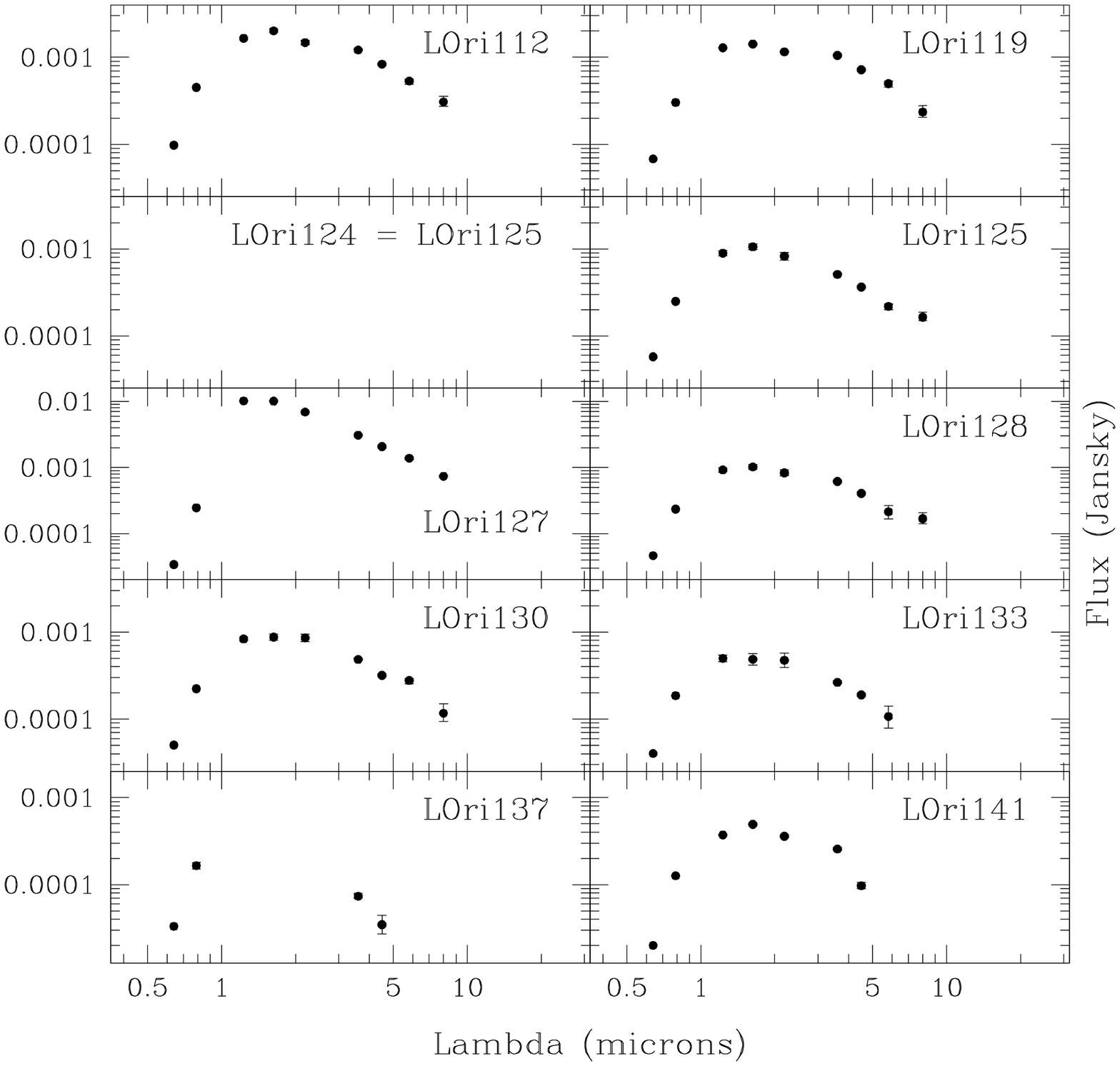}
    \includegraphics[width=4.2cm]{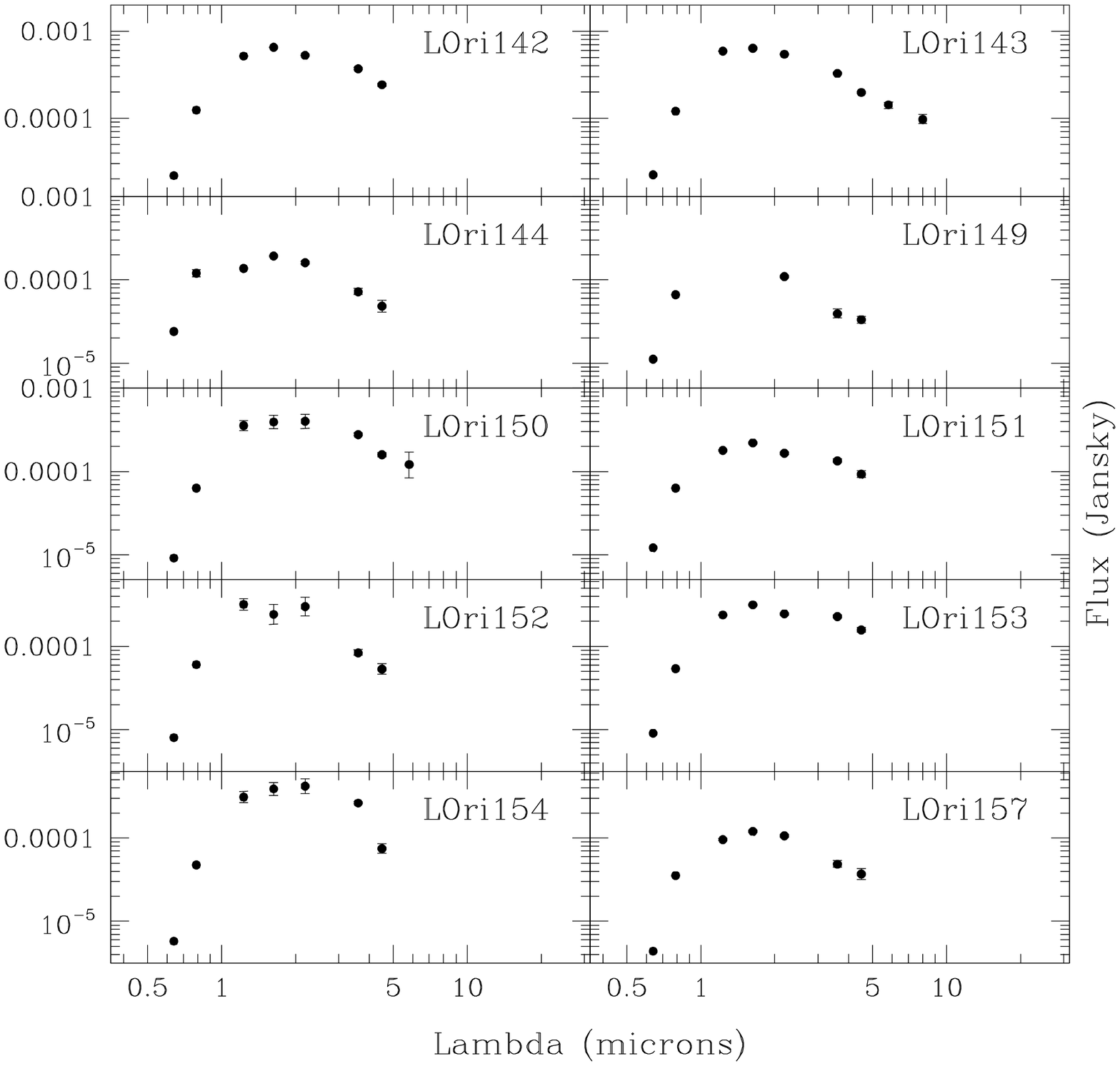}
    \includegraphics[width=4.2cm]{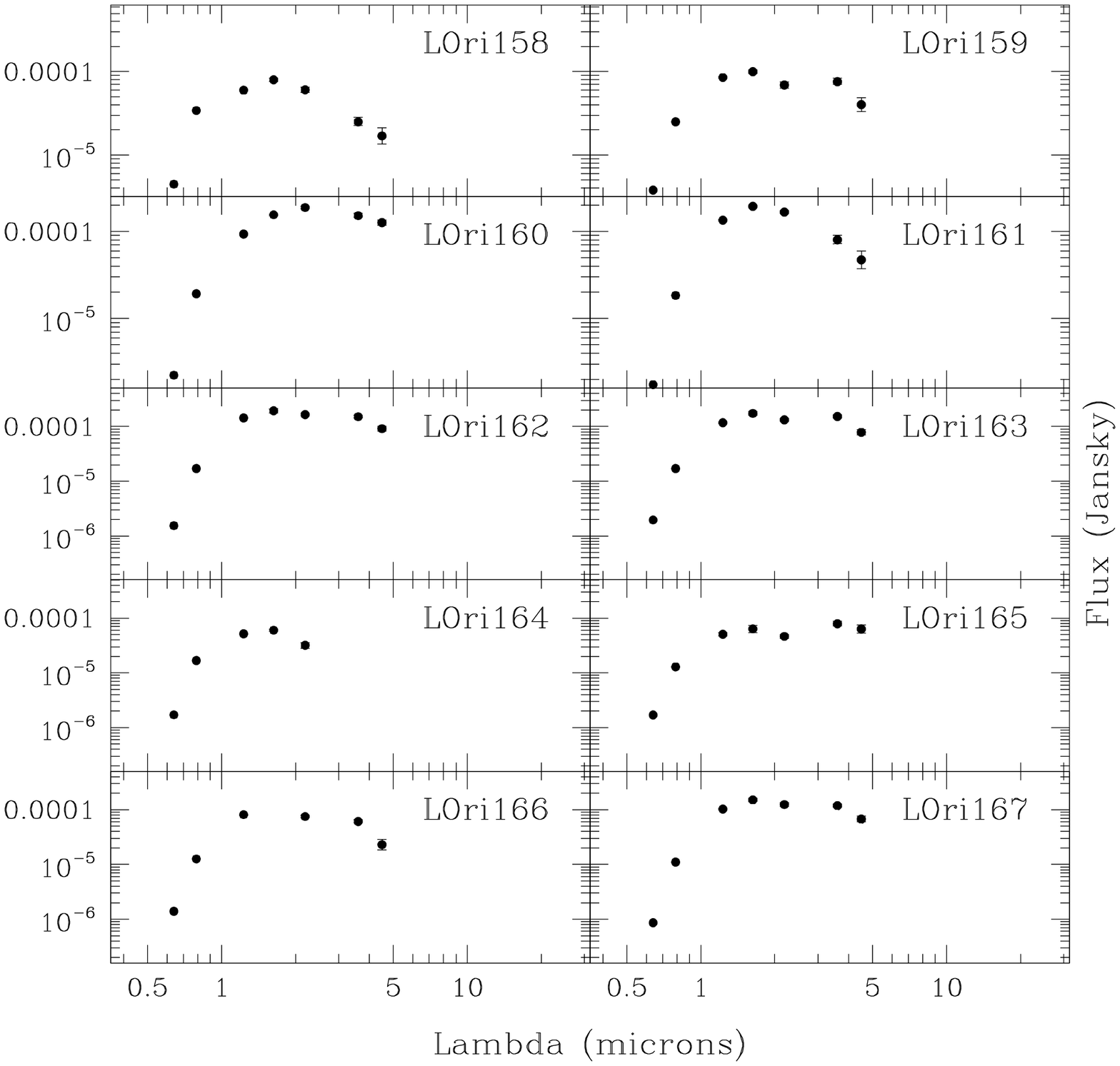}
    \includegraphics[width=4.2cm]{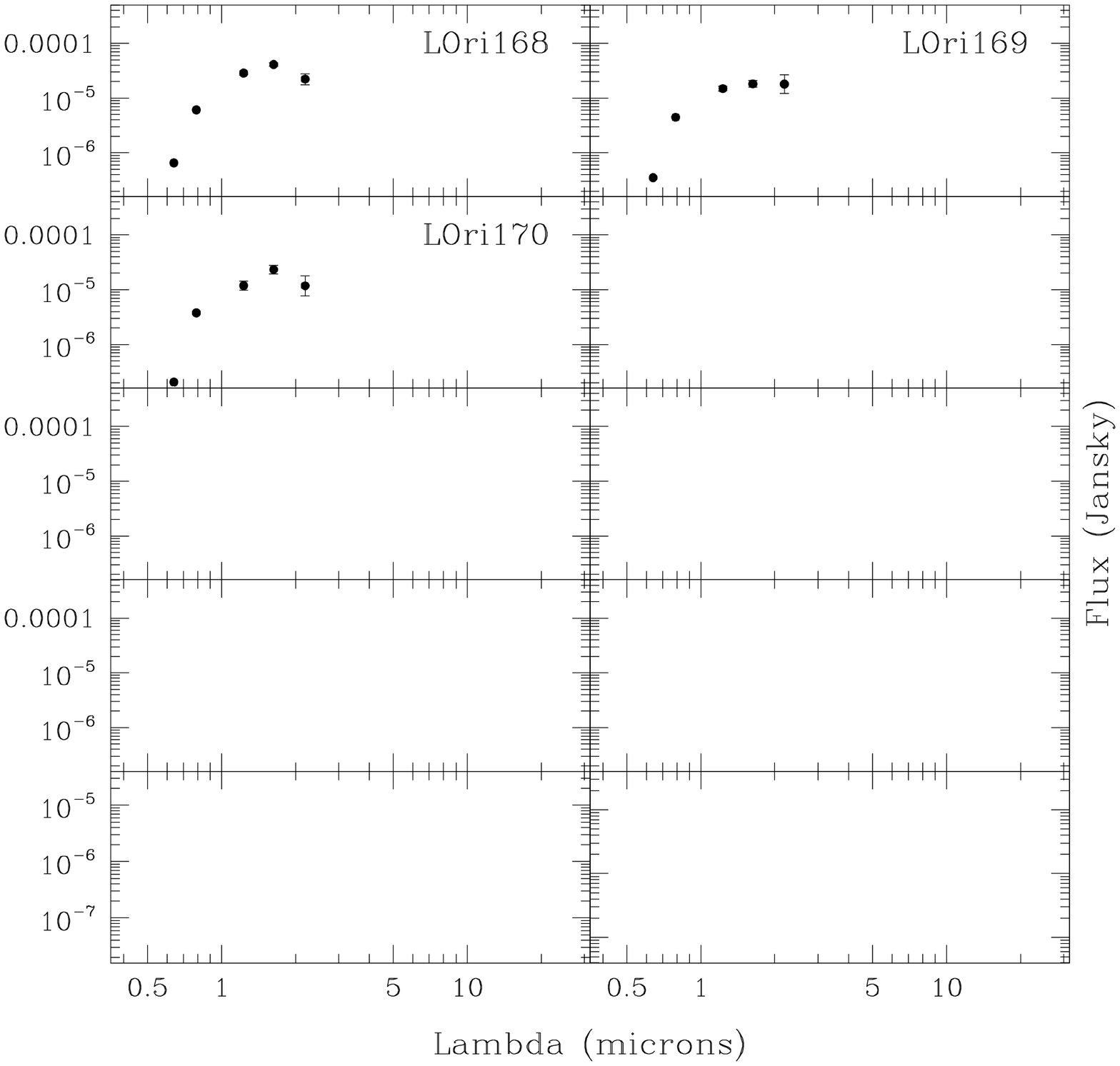}
 \caption{
Spectral Energy Distributions for some stellar members of the Lambda
Orionis cluster sorted according to their IRAC slope: 
simple photosphere spectra. Objects lacking IRAC 
slope or being in the boundary between two types have been classified after visual inspection.
}
 \end{figure*}

\setcounter{figure}{6}
    \begin{figure*}
    \centering
    \includegraphics[width=5.4cm]{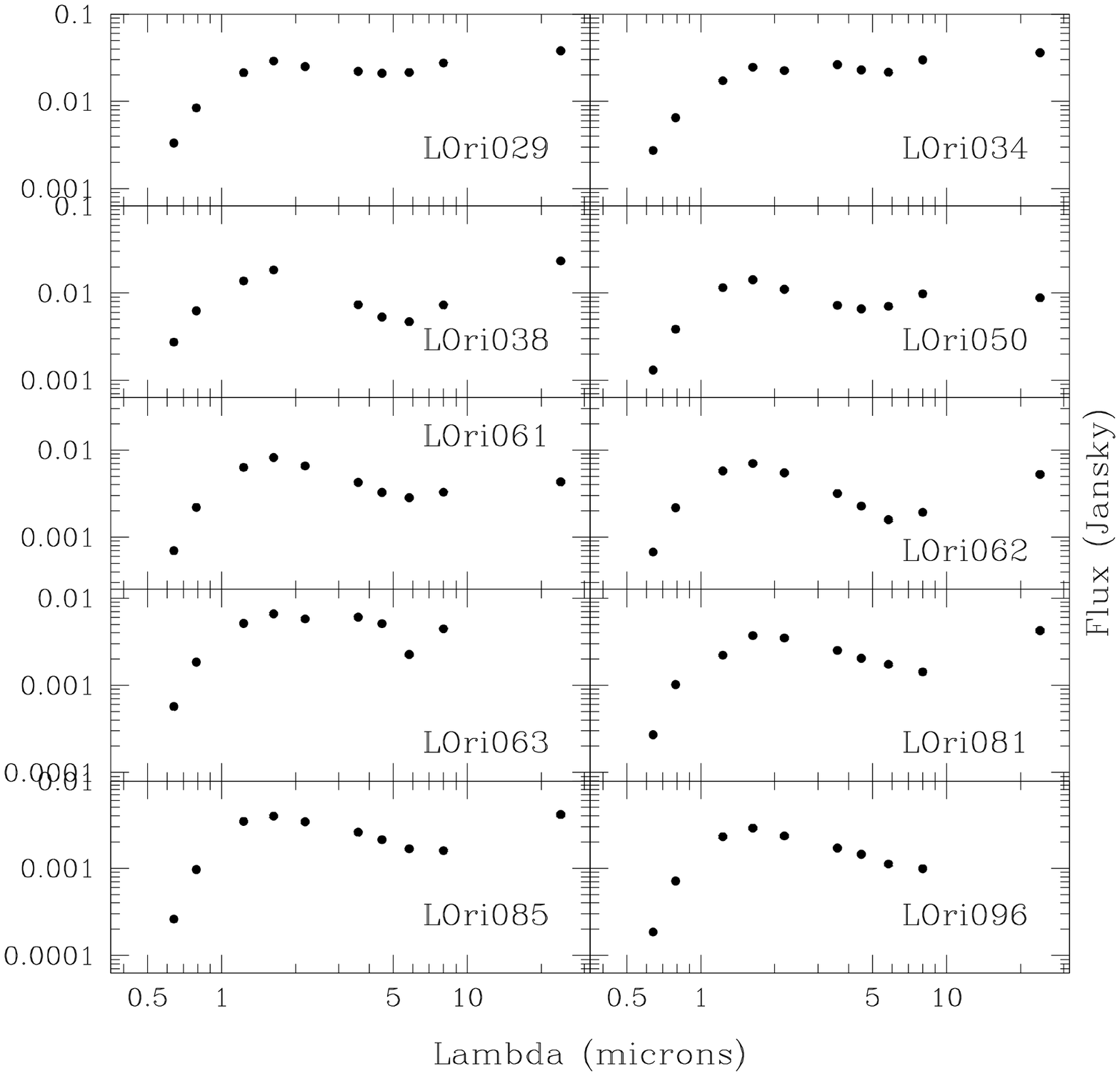}
    \includegraphics[width=5.4cm]{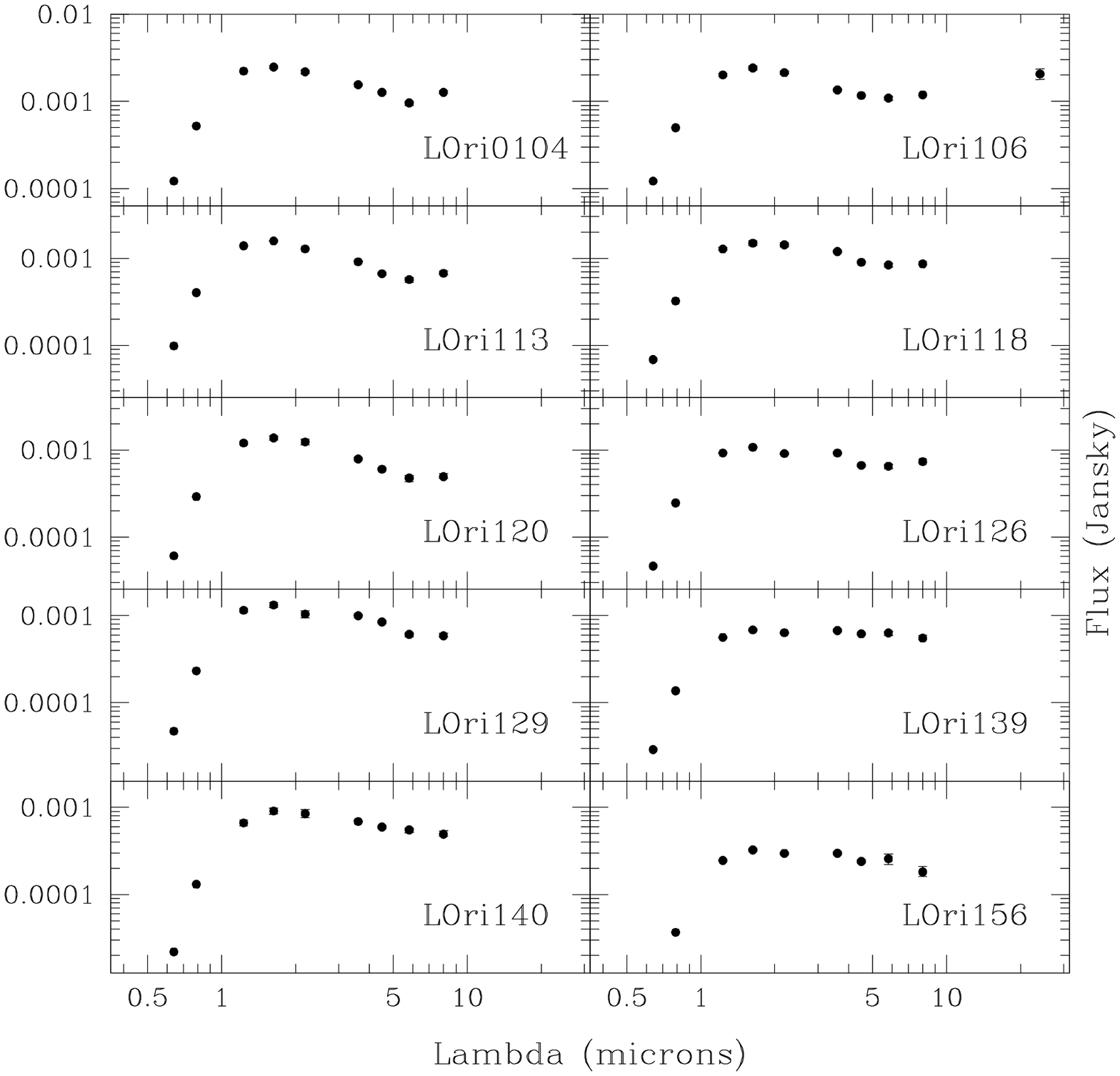}
 \caption{
 Spectral Energy distribution for some stellar members of the Lambda
Orionis cluster sorted according to their IRAC slope: 
flat, or sloping IR spectra with the excess starting in the near-IR (thick disks).
}
 \end{figure*}

\setcounter{figure}{7}
    \begin{figure*}
    \centering
    \includegraphics[width=5.4cm]{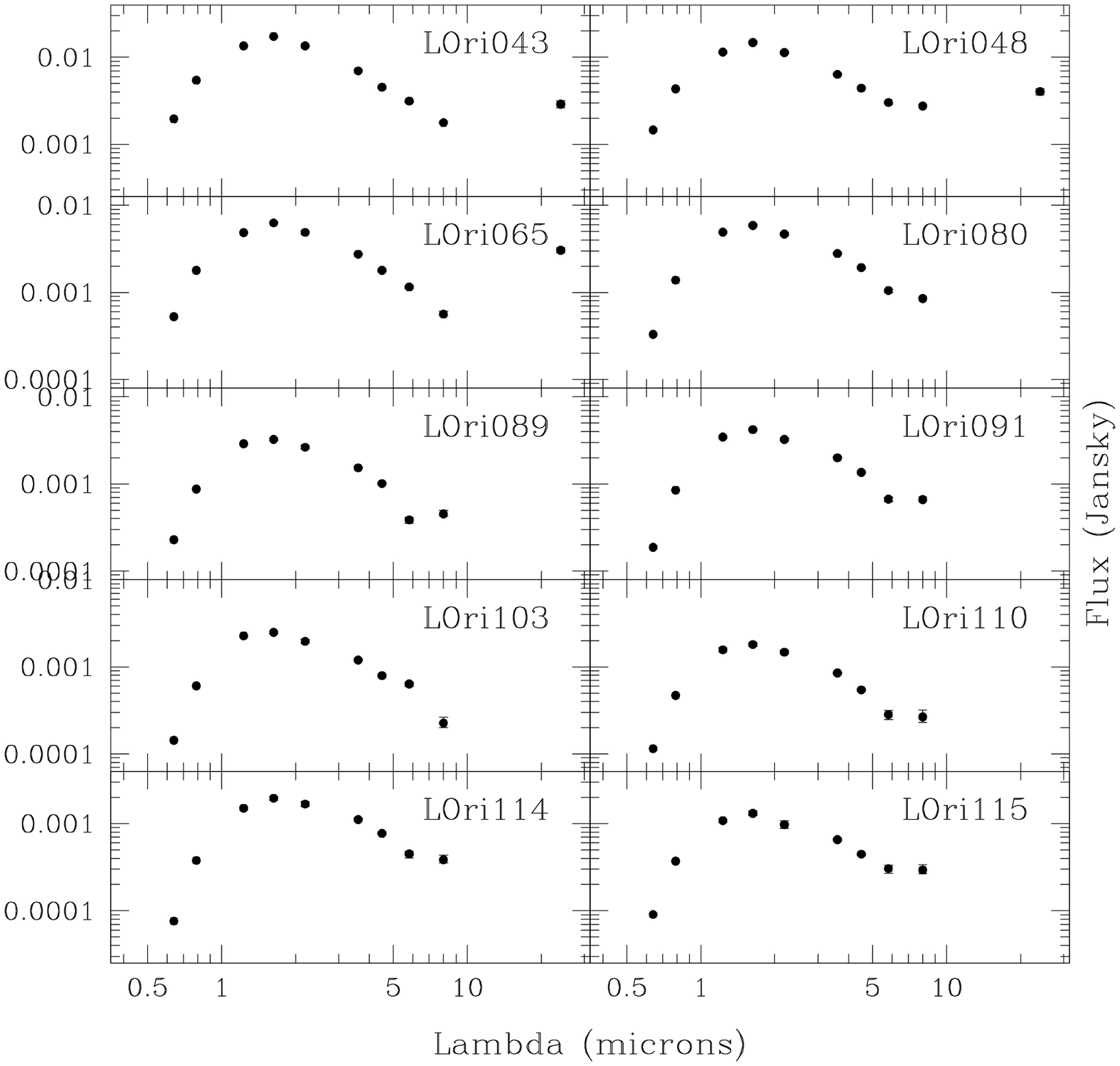}
    \includegraphics[width=5.4cm]{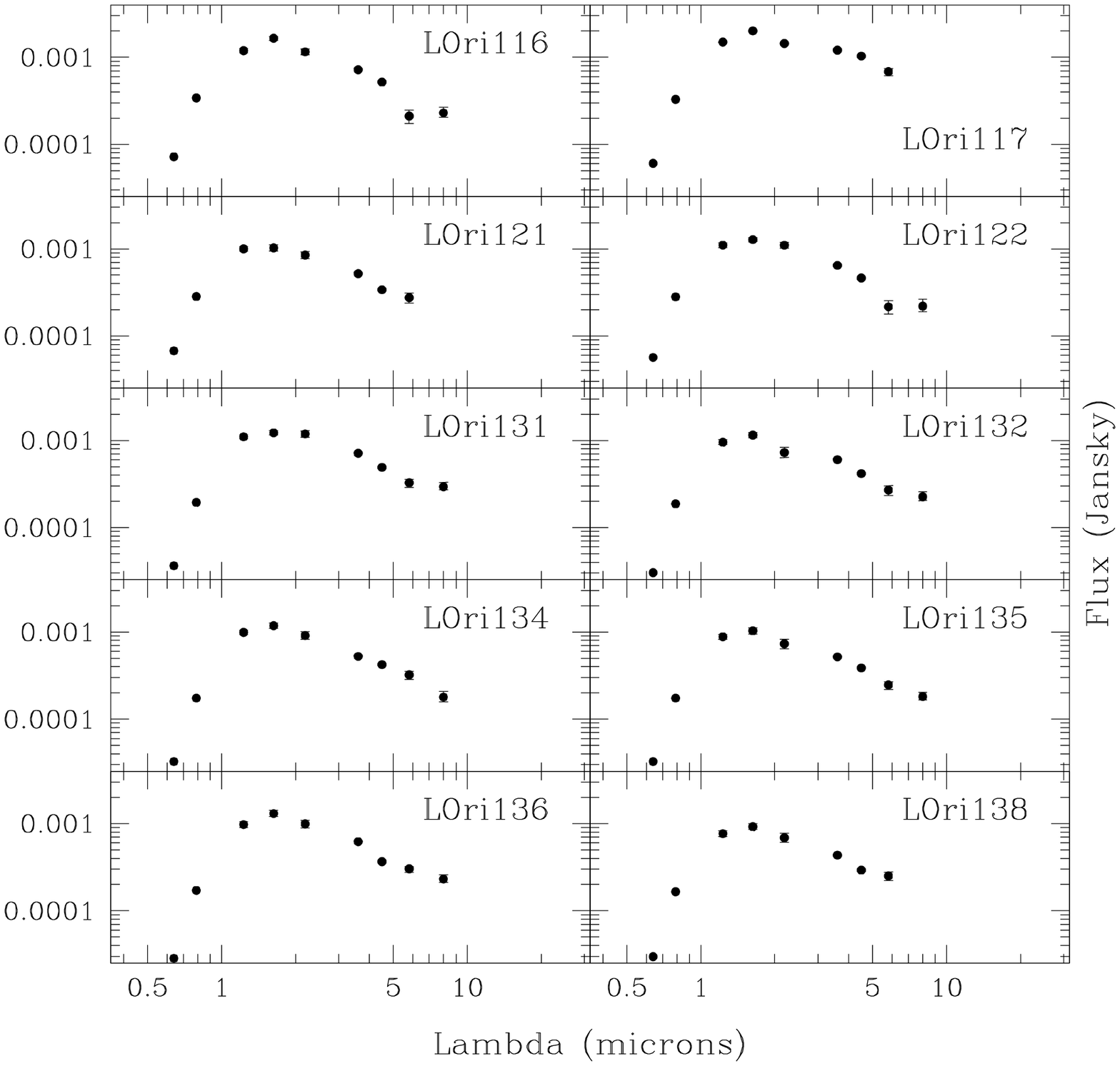}
    \includegraphics[width=5.4cm]{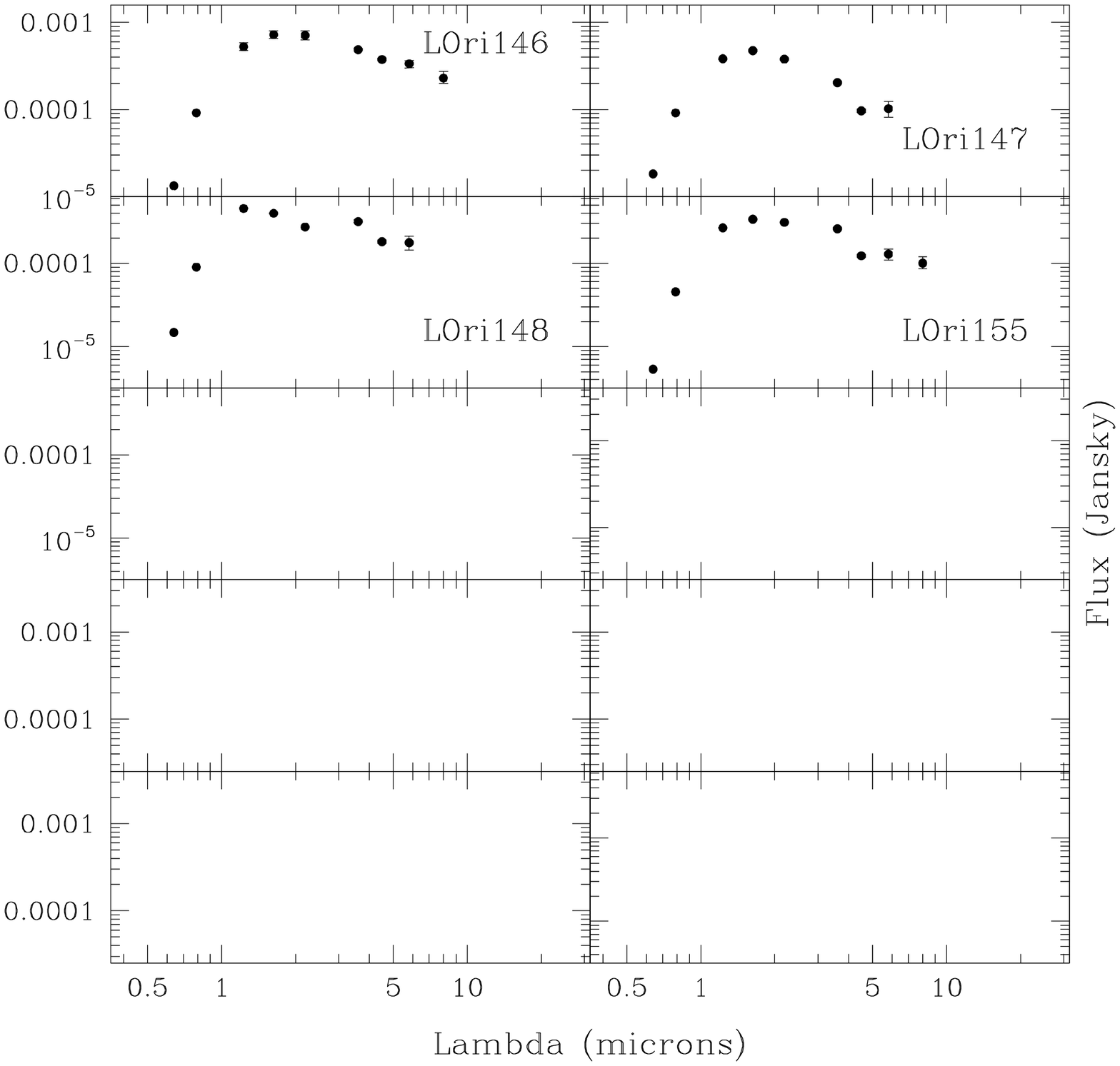}
 \caption{
Spectral Energy distribution for some stellar members of the Lambda
Orionis cluster sorted according to their IRAC slope: 
spectra with excesses begining in the IRAC or MIPS
range (thin disks and transition objects).
 LOri043 and LOri065 were classified as diskless objects 
but have been sorted as objects bearing thin disks due to their excess at MIPS [24].
}
\end{figure*}

\setcounter{figure}{8}
    \begin{figure*}
    \centering
    \includegraphics[width=7.8cm]{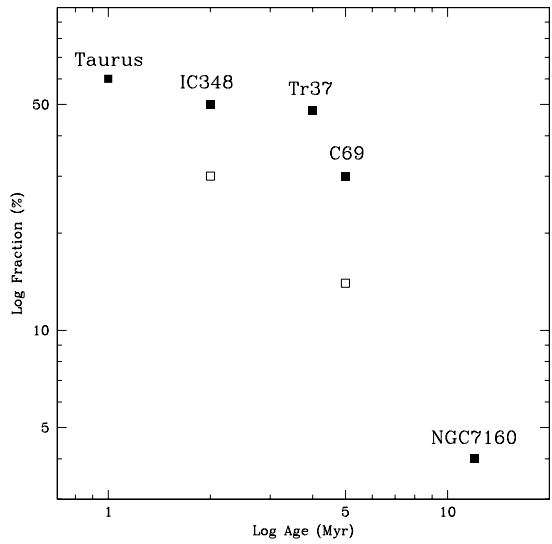}
 \caption{
The fraction of Class~II stars and massive brown dwarfs in several SFRs and young clusters (filled squares). Open squares stand for thick disk fractions of IC348 and C69.
}
 \end{figure*}

\setcounter{figure}{9}
    \begin{figure*}
    \centering
    \includegraphics[width=10.8cm]{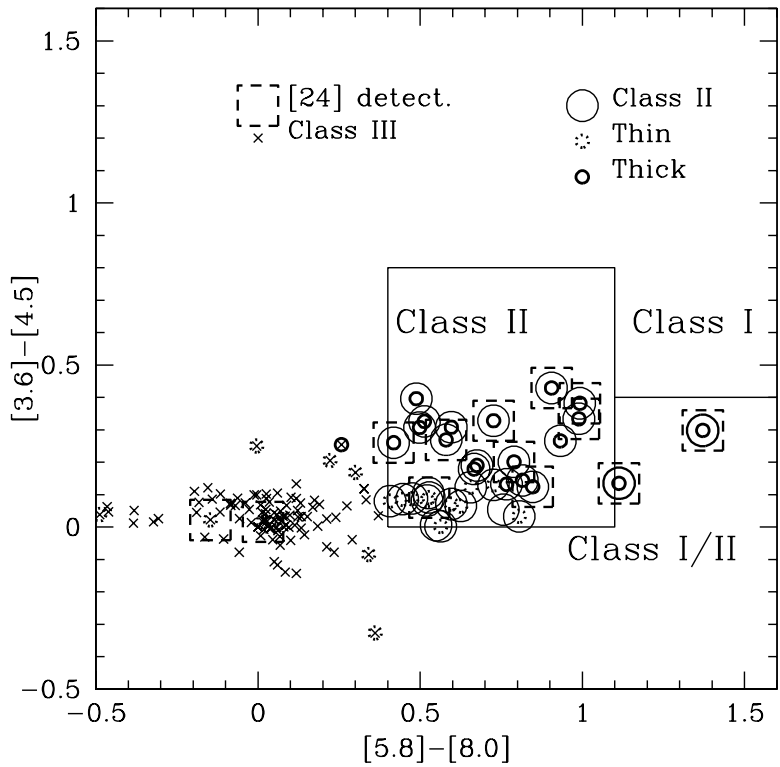}
 \caption{
Spitzer/IRAC CCD. We show with different symbols (see key) 
cluster members with different types of disks. 
}
 \end{figure*}


\setcounter{figure}{10}
    \begin{figure*}
    \centering
    \includegraphics[width=7.8cm]{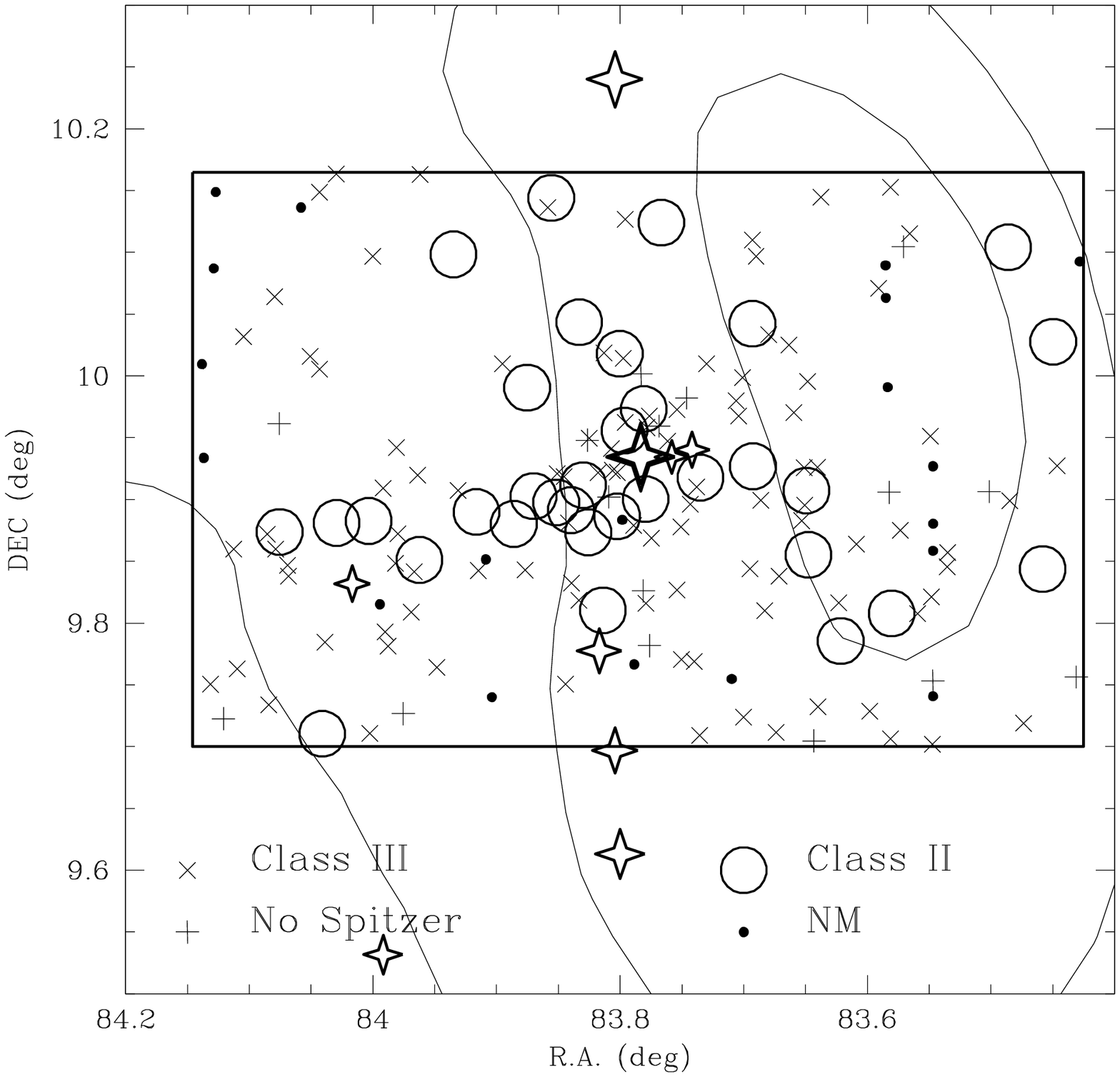}
    \includegraphics[width=7.8cm]{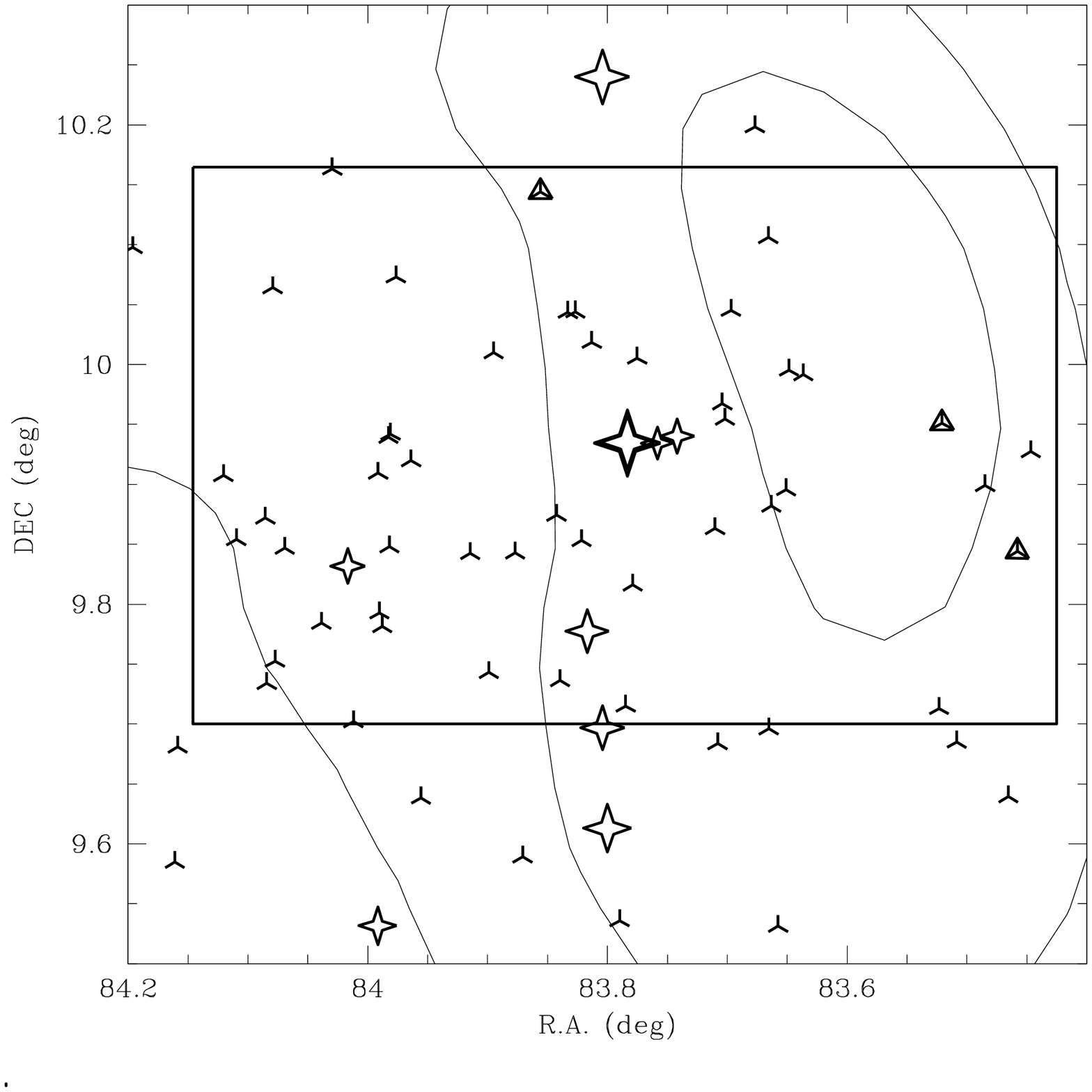}
 \caption{
{\bf a)}
Spatial distribution of our sample.
 IRAS contour levels at 100 microns also have have been 
included as solid (magenta) lines. The big, thick-line rectangle corresponds to the
CFHT
survey (Paper~I).
Class~II sources  (Classical TTauri stars and substellar analogs) 
have been included as big (red) circles, whereas Class~III (Weak-line TTauri) 
objects appear as crosses, and other Lambda Orionis
members lacking the  complete set of IRAC photometry are displayed with the  plus
symbol.
{\bf b)}
Spatial distribution of the low mass stars from Dolan \& Mathieu (1999, 2001).
 OB stars  appear as four-point (blue) stars, with size related to magnitude
 (the bigger, the brighter).
The overplotted thick triangles indicates those stars whose H$\alpha$\ equivalent 
width is larger than the saturation criterion defined by 
Barrado y Navascu\'es \& Mart\'{\i}n (2003), thus suggesting the presence of active
accretion. Based on H$\alpha$  alone, the fraction of accreting stars would be 11\%.}
 \end{figure*}


\setcounter{figure}{11}
    \begin{figure*}
    \centering
    \includegraphics[width=6.7cm]{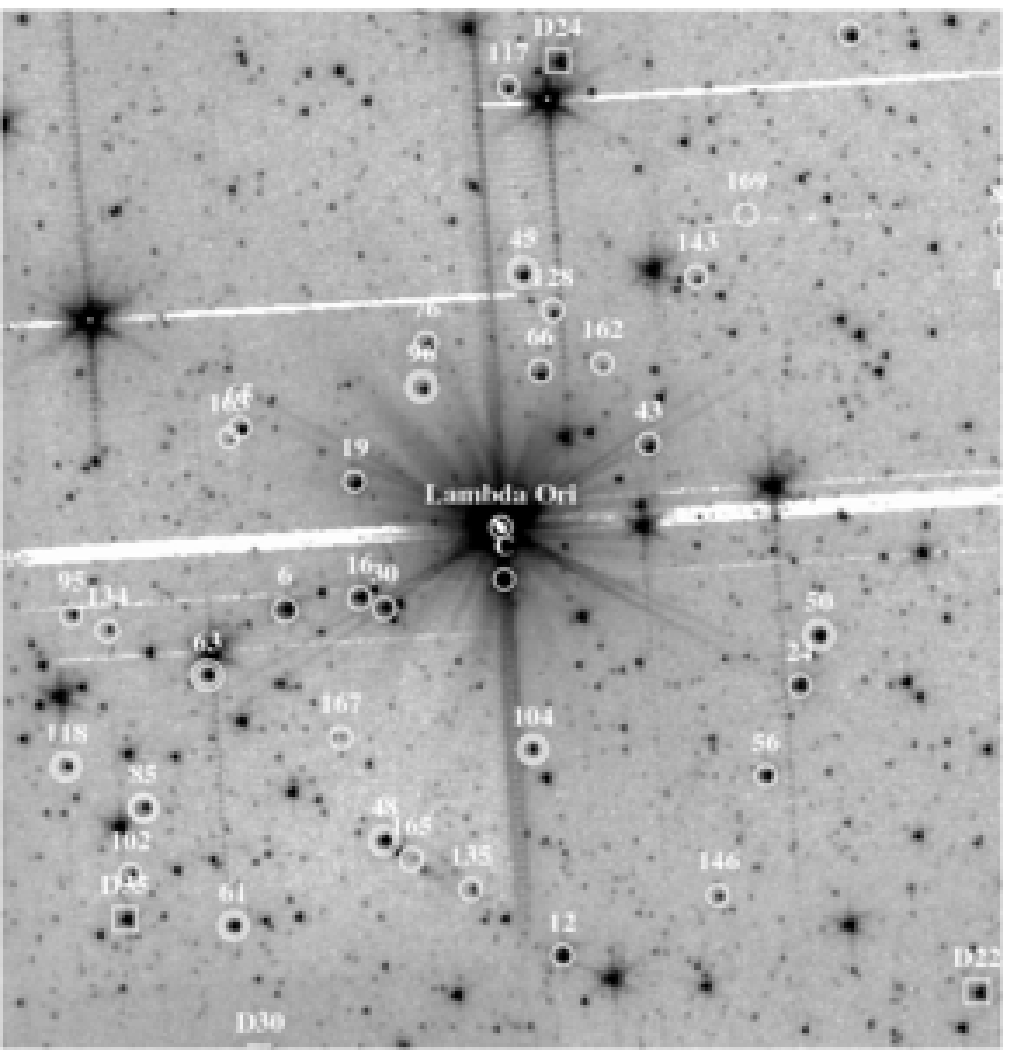}
    \includegraphics[width=6.8cm]{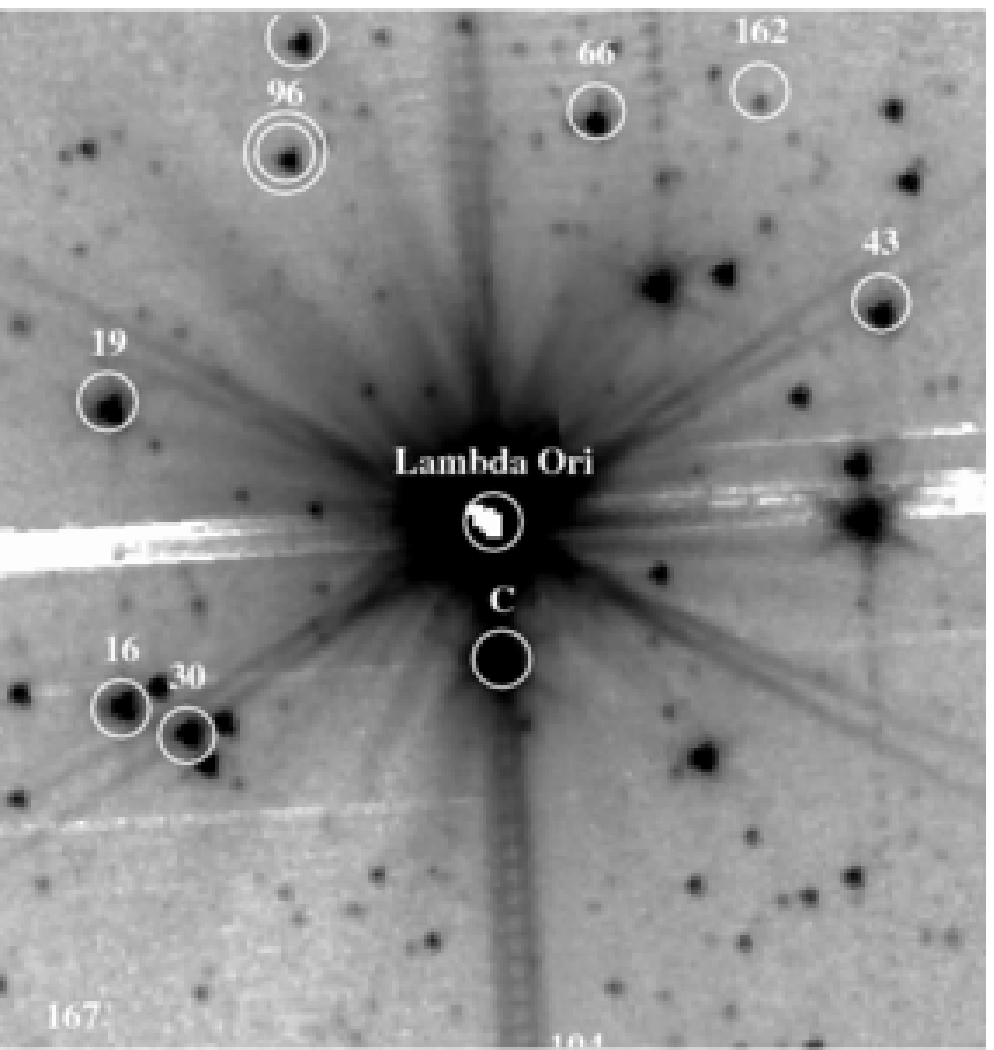}
    \includegraphics[width=11.8cm]{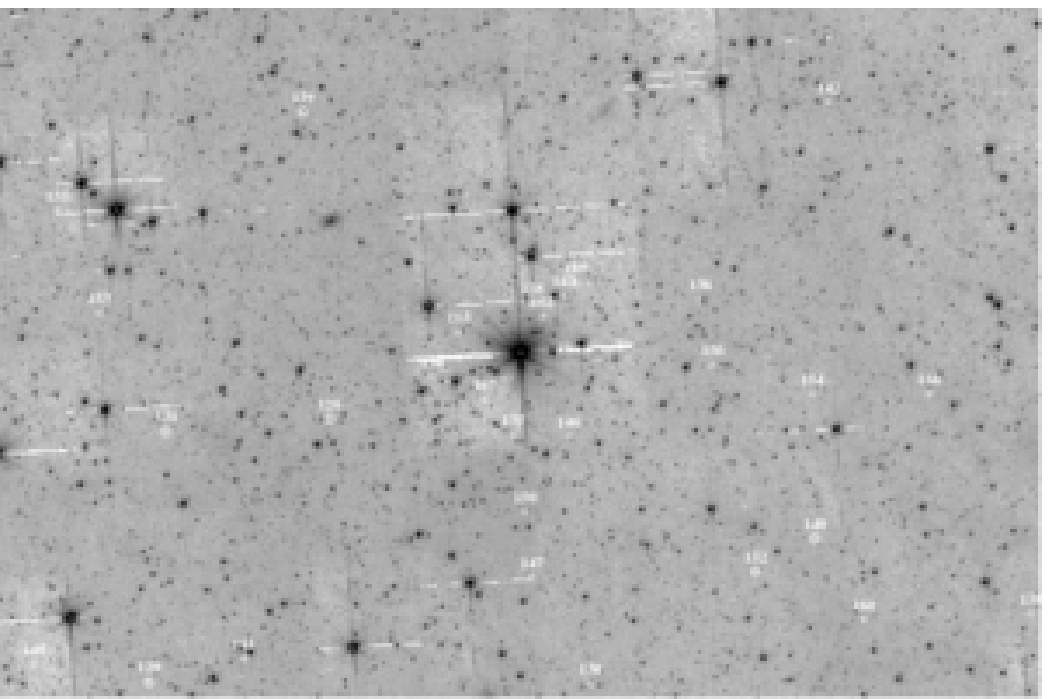}
 \caption{
Spitzer/IRAC image at 3.6 micron centered around the star $\lambda$ Orionis.
{\bf a)}
The size is about 9$\times$9 arcmin, equivalent to 192,000 AU. 
The double circle indicates the presence of a Class~II object, whereas
squares indicate the location of cluster members from Dolan \& Mathieu (1999;2001).
The intensity of the image is in logarithmic scale.
{\bf b)} Detail around the star $\lambda$ Orionis. 
The size is about 3.3$\times$3.3 arcmin, equivalent to 80,000 AU. 
{\bf c)}
Distribution of bona-fide brown dwarfs.
 The size of the image is 45$\times$30 arcmin.
North is up, East is left.
}
 \end{figure*}


\setcounter{figure}{12}
    \begin{figure*}
    \centering
    \includegraphics[width=16.8cm]{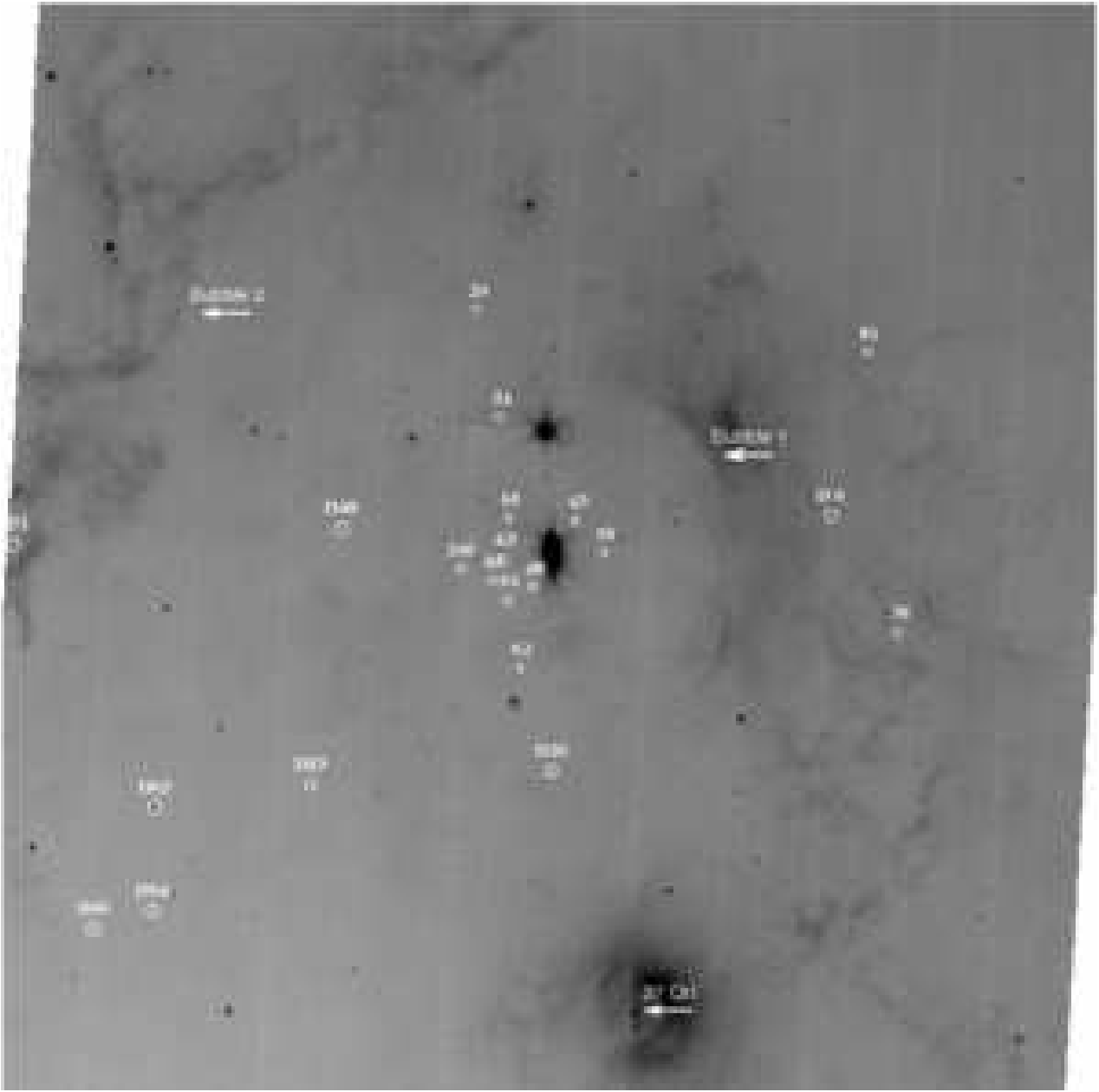}
 \caption{
Spitzer/MIPS image at 24 microns which includes the members of the Lambda Orionis
cluster
visible at this wavelength,  including those cluster members 
by Dolan \& Mathieu (1999, 2001)
as big circles and CFHT member as small circles detected at this wavelength.
 The size is about 60.5$\times$60.5 arcmin. 
North is up, East is left. 
The figure is centered on the star $\lambda$ Ori AB.}
 \end{figure*}




\end{document}